\documentclass[aps,prl,showpacs,showkeys,twocolumn,superscriptaddress,
              floats,floatfix,preprintnumbers]{revtex4-1}
\usepackage{epsfig,graphicx,amsmath,amssymb}
\usepackage{mathrsfs}
\usepackage{bm}
\usepackage{color}
\def \rhor {\rho_{\rm r}}

\def \AA {\mathcal{A}}
\def \BB {\mathcal{B}}
\def \ez {\hat{e}_z}

\def \PC {P^{\rm C}}
\def \Pr {P^{\rho}}
\def \Qr {Q^{\rho}}
\def \PE {P^{\rm E}}
\def \PL {P^{\rm L}}

\def \ap {a_{\rm p}}

\def \TA {T_{\rm A}}

\def \rhof {\rho_{\rm f}}
\def \rhop {\rho_{\rm p}}

\def \uu  {{\bm u}}
\def \vv  {{\bm v}}

\def \teta {t_{\rm \eta}}

\def  \XX  {{\bm X}}
\def \Xdot {\dot{{\bm X}}}
\def \Vdot {\dot{{\bm V}}}
\def  \VV  {{\bm V}}
\def \Vdot {\dot{{\bm V}}}

\def \taup {\tau_{\rm p}}

\def \grad {{\bm \nabla}}
\def \curl {{\bm \nabla} \times}
\def \dive {{\bm \nabla}\cdot}
\def \lap {\nabla^2}
\def \delt {\partial_t}

\newcommand{\bra}[1]{\langle #1\rangle}
\def \Rey  {\mbox{Re}}

\def \St  {\mbox{St}}

\def \kf  {k_{\rm f}}

\def \Np  {N_{\rm p}}

\def \Erho {E_{\rm \rho}}
\def \Ev   {E_{\rm v}}

\def \PL   {P^{\rm L}}
\def \PEd  {P^{\rm \rho}}
\def \TA   {\Tr{\AA}}
\def \DA   {\Det{\AA}}
\def \DB   {\Det{\BB}}

\newcommand{\beq}{\begin{equation}}
\newcommand{\eeq}{\end{equation}}
\DeclareMathOperator{\Tr}{Tr}
\DeclareMathOperator{\Det}{Det}

\newcommand{\REM}[1]{{}}

\newcommand{\Eq}[1]{Eq.~(\ref{#1})}
\newcommand{\Fig}[1]{Fig.~(\ref{#1})}
\newcommand{\subfig}[2]{Fig.~(\ref{#1}#2)}

\usepackage{tgtermes}

\begin{document}
\title{Topology of two-dimensional turbulent flows of dust and gas} 
\author{Dhrubaditya Mitra}
\email{dhruba.mitra@gmail.com}
\affiliation{Nordita, KTH Royal Institute of Technology and
Stockholm University, Roslagstullsbacken 23, 10691 Stockholm, Sweden}
\author{Prasad Perlekar} 
\email{perlekar@tifrh.res.in}
\affiliation{Tata Institute of Fundamental Research,  Centre for Interdisciplinary Sciences, Hyderabad-500107, India.}
\begin{abstract}
We perform direct numerical simulations (DNS) of passive heavy inertial particles (dust) in 
homogeneous and isotropic two-dimensional turbulent flows (gas) for a range of Stokes 
number, $\St < 1$.  
We solve for the particles using both a Lagrangian and an Eulerian approach (with a 
shock-capturing scheme). In the latter the particles are described by a 
dust-density field and a dust-velocity field. 
We find that: 
The dust-density field in our Eulerian simulations have the same correlation 
dimension $d_2$ as obtained from the clustering of particles in the Lagrangian 
simulations for $\St < 1$;
the cumulative probability distribution function (CDF) of the dust-density coarse-grained over
a scale $r$ in the inertial range has a left-tail with a power-law fall-off
indicating presence of voids; 
The energy spectrum of the dust-velocity has a power-law range with an exponent
that is same as the gas-velocity spectrum except at very high Fourier modes;
The compressibility of the dust-velocity field is proportional to $\St^2$.
We quantify the topological properties of the dust-velocity and the gas-velocity
through their gradient matrices, called $\AA$ and $\BB$, respectively.
Our DNS confirms that the statistics of topological properties of $\BB$ are 
the same in Eulerian and Lagrangian frames only if the Eulerian data are weighed 
by the dust-density.
We use this correspondence to study the statistics of topological properties of
$\AA$ in the Lagrangian frame from our Eulerian simulations by 
calculating density-weighed probability density functions. 
We further find that in the Lagrangian frame the mean value of the trace of $\AA$, 
$\bra{\Tr\AA}^{\rho}$ is negative and its magnitude increases with $\St$ 
approximately as $\exp(-C/\St)$ with a constant $C\approx 0.1$.
The statistical distribution of different topological structures that appear in the
dust flow are different in Eulerian and Lagrangian (density-weighed Eulerian) cases 
particularly for $\St$ close to unity. 
In both of these cases, for small $\St$ the topological structures
have close to zero divergence and are either vortical (elliptic) or
strain-dominated (hyperbolic, saddle).  As $\St$ increases, the contribution 
to negative divergence comes mostly from saddles and the contribution to positive 
divergence comes from both vortices and saddles. Compared to the Eulerian case, the 
density-weighed Eulerian case has less inward spirals and more converging saddles. 
Outward spirals are the least probable topological structures in both cases.
\end{abstract}
\preprint{NORDITA 2017-115}
\maketitle
\section{Introduction}
\label{intro}
Particle laden flows appear in a variety of natural phenomena such as transport of aerosol, 
volcanic ash in atmosphere~\cite{kok2012physics}, 
raindrop formation~\cite{Pruppacher2010microphysics}, 
and formation of planetesimals in propoplanetary disks~\cite{Arm10}.  
To describe such multiphase flows, in what follows, we shall call the particles ``dust'' phase 
and the carrier fluid ``gas'' phase.  
In many cases the suspension is dilute enough to ignore 
inter-particle interactions. The dynamics of particles in this limit is controlled by their 
size and density $\rhop$ relative to that of the carrier fluid $\rhof$. 
For small particles -- smaller than the dissipative scale of the flow -- 
with $\rhop \gg \rhof$ (heavy particle limit), the equations of motion further 
simplify and the particle dynamics is determined by a single relaxation 
time $\taup$, such that the equation of motion of a single particle is given by 
\begin{subequations}
\begin{align}
\Xdot &= \VV \/, \label{eq:dxdt}\\
\Vdot &= \frac{1}{\taup}\left[ \uu(\XX) - \VV \right] \/.
\label{eq:dvdt}
\end{align}
\label{eq:HIP}
\end{subequations}
Here the dot denotes time differentiation, $\XX$ and $\VV$ are
respectively the position and velocity of a particle, and $\uu$ is the flow velocity at 
the particle position that is determined by solving the Navier--Stokes equation with 
appropriate boundary conditions.
Such particles are called heavy inertial particles (HIPs). 

A crucial quantity in this subject is the  (binary) collision rate -- mean number of collisions 
between dust particles per-unit-volume per-unit-time and also the probability density 
function~(PDF) of their collisional velocities.
In most cases of practical interests, the effects of gravity (for example, leading to sedimentation
 of dust) is important because differently sized dust particles reach different terminal speeds 
under gravity; this can be the dominant term in the collision rate. 
But our primary interest is the effect of turbulence on the collision rate, 
hence gravity is ignored in the rest of this paper.
Furthermore, we consider all the dust particles to have the same size (monodisperse suspension), 
hence the differential terminal speeds under gravity are not relevant.
The collision rate is then proportional to the mean relative velocity of two dust particles 
separated by a distance of $2\ap$, multiplied by the probabilty of finding two dust particles 
with separation $2\ap$ where $\ap$ is the radius of the particles.
Let us first consider the latter, v.i.z., the probability of finding two dust particles at 
a separation less than  $R$; $P_2(R)$.  
If the particles are homogeneously distributed then $P_2(R) \sim R^{d}$ where $d$ 
is the dimension of space. 
Direct numerical simulations (DNS) in the last two decades have conclusively 
shown~\cite{bec07} that in homogeneous and isotropic turbulence the heavy inertial 
particles show clustering, i.e., 
$P_2(R) \sim R^{d_2}$ with 
$d_2 \leq d$, a non-monotonic function of the Stokes number, 
$\St\equiv \taup/\teta$, where $\teta$ is the Kolmogorov time. 
Clearly this clustering increases the collision rate. 
Next consider the former, mean relative velocity of two dust particles separated by a distance $R$. 
On one hand, if the dust-velocity field is smooth and $\ap$ is small enough, the mean relative 
velocity is determined by the mean value of the gradient of dust-velocity field~\cite{saf+tur56}. 
On the other hand, if the dust-velocity field develops discontinuties, e.g., shocks, then 
the mean value of jumps across the discontinuties and the number density of the 
discontinuties contribute to the collision rate. 
Either way, it is the gradient matrix of the dust-velocity field, $\AA$,  that plays a central role.
It can be shown~\cite{fal02,wilkinson2005caustics} that $\AA$ obeys the equation:
\begin{equation}
\dot{A}_{ij} = \frac{1}{\taup}\left[B_{ij}-A_{ij} \right] - A_{ik}A_{kj}
\label{eq:adot}
\end{equation}
in Lagrangian frame, i.e., when \Eq{eq:adot} is solved along the trajectory of a dust particle. 
Here we have assumed Einstein convention of summing of repeated indices. 
The matrix $\BB$ with component $B_{ij}$ is the gradient-matrix of the gas-velocity, i.e., 
$B_{ij} \equiv \partial_j u_i$.
It is now well-established~\cite{fal02,wilkinson2005caustics,wilkinson2006caustic,
gustavsson2014relative} that under quite general conditions \Eq{eq:adot} may develop singularities 
in finite-time -- called caustics.
In turbulent flows  $\BB$ and consequently $\AA$ are random matrices that are not necessarily 
Gaussian or white-in-time. 
Hence we must take a statistical approach in trying to understand them.  
To the best of our knowledge, except Ref.~\cite{falkovich2007sling}, there has been no attempt to 
calculate the statistical properties of the matrix $\AA$ in turbulent flows.
It has been calculated analytically~\cite{wilkinson2006caustic} in one dimensional models 
where $\AA$ and $\BB$ are reduced to real variables (instead of matrices) and 
$\BB$ is assumed to be a Gaussian, white-in-time, process. 
In this paper our aim is to study both $\BB$ and $\AA$ in turbulent 
flows in two spatial dimensions. 

Here we limit ourselves to incompressible two-dimensional ($2D$) turbulent flows that are 
considered to be a minimalistic model for atmospheric and oceanic 
flows~\cite{bof12,dani00,sch02,per09b,pan17}. 
Two-dimensional turbulence is characterised by an inverse cascade of energy to length scales 
larger than the forcing scale and a forward cascade of enstrophy from forcing 
scale down to the small scales~\cite{kraic67,lei68,bat69}. 
In general, local topological properties of a two-dimensional flow can be characterised by the 
trace, $\Tr$, and the determinant, $\Det$, of the velocity gradient matrix \cite{oku70}. 
For an incompressible flow of gas, $\Tr\BB=0$. 
Hence the flow consists of only elliptic (vortical) or saddle (strain dominated) points which are 
characterised in the terms of the sign of $\Lambda \equiv \Det \BB$ (also known as the Okubo-Weiss 
criterion \cite{oku70,wei92}). 
A statistical description of the topology of the flow is given by the PDF of $\DB$ alone. 
Soap-film experiments \cite{riv01} and numerical studies of 
Navier-Stokes equations~\cite{bas94,per09} have used Okubo-Weiss criterion to investigate 
the topological properties of incompressible two-dimensional turbulent flows. 
In the light of the discussion of the previous paragraph, it is clear that the 
dust-velocity is compressible even when the gas-velocity is incompressible.  
Hence a statistical description of the topological structures found in 
the dust-flow is given by the joint PDF of $\Det\AA$ and $\Tr\AA$. 
This is the principal objective of this paper. 

More specifically, we investigate the topological properties of the dust-velocity field 
passively advected by the forward cascade of an incompressible two-dimensional turbulent flow. 
In principle, either Eulerian or Lagrangian framework~(see Ref.~\cite{eat10} for an overview)
can be used in direct numerical simulations~(DNS) to calculate the gradient
matrices $\AA$ and $\BB$. 
In the Eulerian framework, mass and momentum 
balance equations are written for both the gas phase as well as the dust phase. 
It is straightforward to calculate the gradient matrices $\AA$ and $\BB$ as both the 
gas-velocity and the dust-velocity field are available on a grid. 
But, note that \Eq{eq:adot} is formulated in the Lagrangian frame, and there is no 
a-priori reason to assume that statistical properties of $\BB$ or $\AA$ in the two 
frames are necessarily the same. 
In the Lagrangian framework the equations of motion for the gas phase  
remains the same but the individual inertial particle trajectories are evolved.
It is straightforward to calculate $\BB$ at the instantaneous position of a dust particle 
by interpolation, but it is non-trivial to calculate $\AA$ -- one must 
solve \Eq{eq:adot} on each dust particle~\cite{falkovich2007sling}. 
In either Lagrangian or Eulerian framework the possibility of development of singularities 
in $\AA$ poses a numerical problem. 
In our Eulerian DNS we regularize this singularity by a shock-capturing 
scheme.

The rest of the paper is organized as follows.
In section~\ref{model} we describe our model including a summary of the numerical techniques 
we have used. In section~\ref{results} we  show that :
(a) The dust-density field in our Eulerian simulations have the same correlation dimension 
$d_2$ as obtained from the clustering of particles in the Lagrangian simulations for $\St < 1$. 
(b) The dust-density coarse grained over a scale $r$ in the inertial range, $\rhor$,
calculated from our Eulerian DNS shows large fluctuations. 
We quantify these fluctuations by computing the cumulative probability 
distribution function (CDF), $\PC(\rhor)$. This CDF has a left-tail with power-law fall-off
that indicates presence of voids in the dust-density.
(c) The energy spectrum of the dust-velocity has a power-law range with an exponent 
that is same as the gas-velocity spectrum except at very high Fourier modes. 
The spectrum of dust-density also shows a scaling range with an exponent of $-1$.
The compressibility of dust velocity field is proportional to $\St^2$.   
(d) The statistics of topological properties of $\BB$ are the same in Eulerian and Lagrangian 
frames only if the Eulerian data are weighed by the dust-density. 
We use this correspondence to calculate statistics of topological properties of
$\AA$ in the Lagrangian frame from our Eulerian simulations by 
calculating density-weighed averages probability density functions.
In particular, we find that: 
(e) The mean value of $\Tr\AA$ in the Lagrangian frame, $\bra{\Tr\AA}^{\rho}$, is negative 
and its magnitude increases with $\St$ approximately as $\exp(-C/\St)$ with a 
constant $C\approx 0.1$.
(f) For small $\taup$,  $\Det\AA \approx \Det\BB$ and  $\Tr\AA\approx 2\taup\Det\BB$. 
(g) The mean value of the PDF density-weighed $\Det\AA$, $\bra{\Tr\AA}^{\rho}$
is negative and its magnitude increases with $\St$ approximately as $\exp(-C/\St)$
with a constant $C\approx 0.1$. 
(h) The statistical distribution of different topological structures that appear in the
dust flow are different in Eulerian and density-weighed Eulerian cases particularly for
$\St$ close to unity. In both of these cases, for small $\St$ the topological structures 
have close to zero divergence and are either vortical (elliptic) or 
strain-dominated (hyperbolic, saddle). 
As $\St$ increases, the contribution to negative divergence comes mostly from 
saddles and the contribution to positive divergence comes from vortices 
and saddles. Compared to the Eulerian case, the density-weighed Eulerian case has
less inward spirals and more converging saddles. Outward spirals are the least 
probable topological structures in both cases.     
We conclude in section~\ref{conc}. 

Before we conclude this introductory section let us note that the field of dust-gas turbulence 
is vast. We have so far touched only those aspects of recent research that are closely 
related to our specific topic of interest.
For a wider introduction to this subject we suggest several recent 
reviews~\cite{tos09,per09b,eat10,pumir2016collisional,gus+meh16} and the references therein. 

\section{Model}
\label{model}
\begin{figure*}[!ht]
\includegraphics[width=0.33\linewidth]{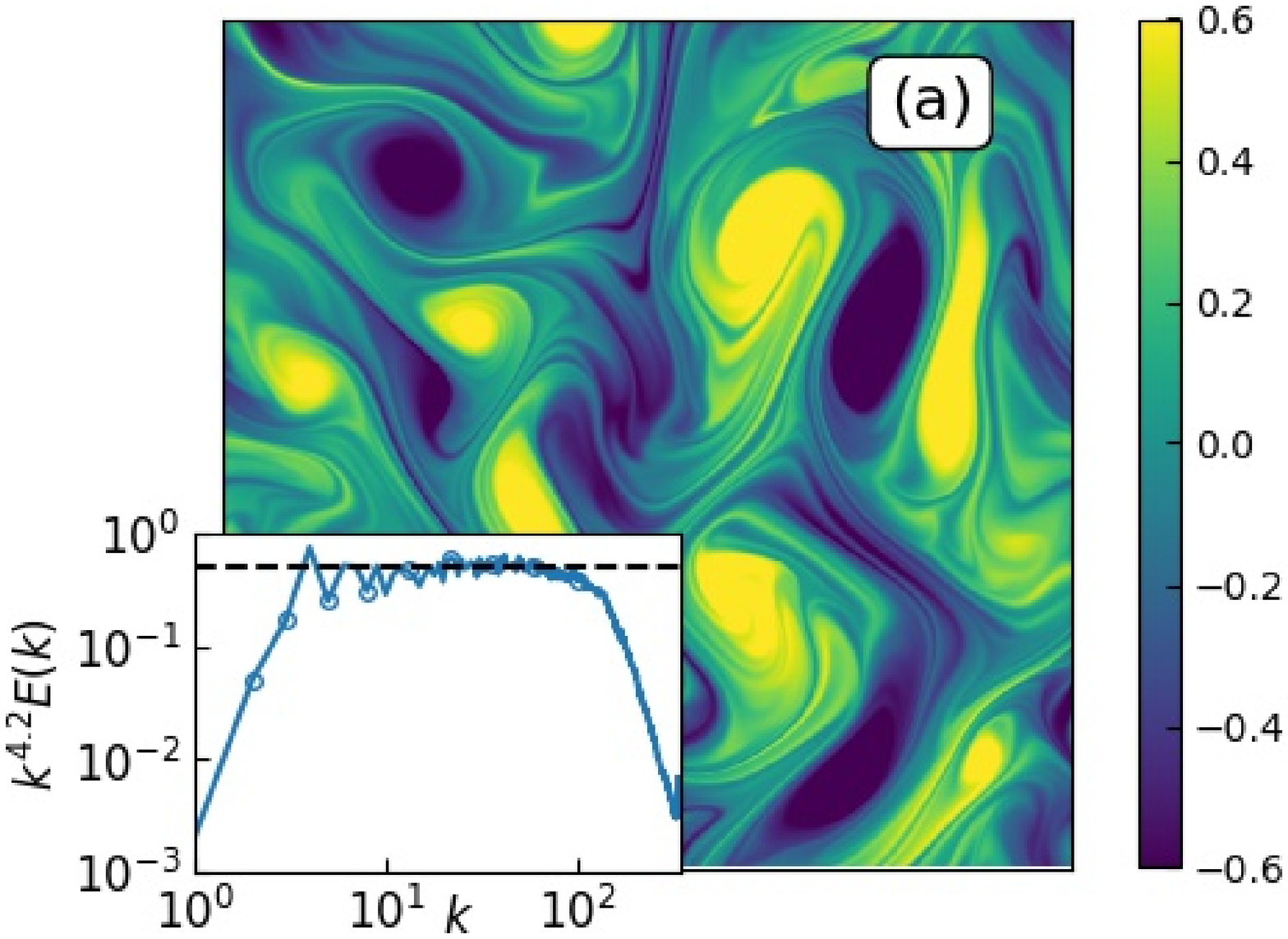} 
\includegraphics[width=0.33\linewidth]{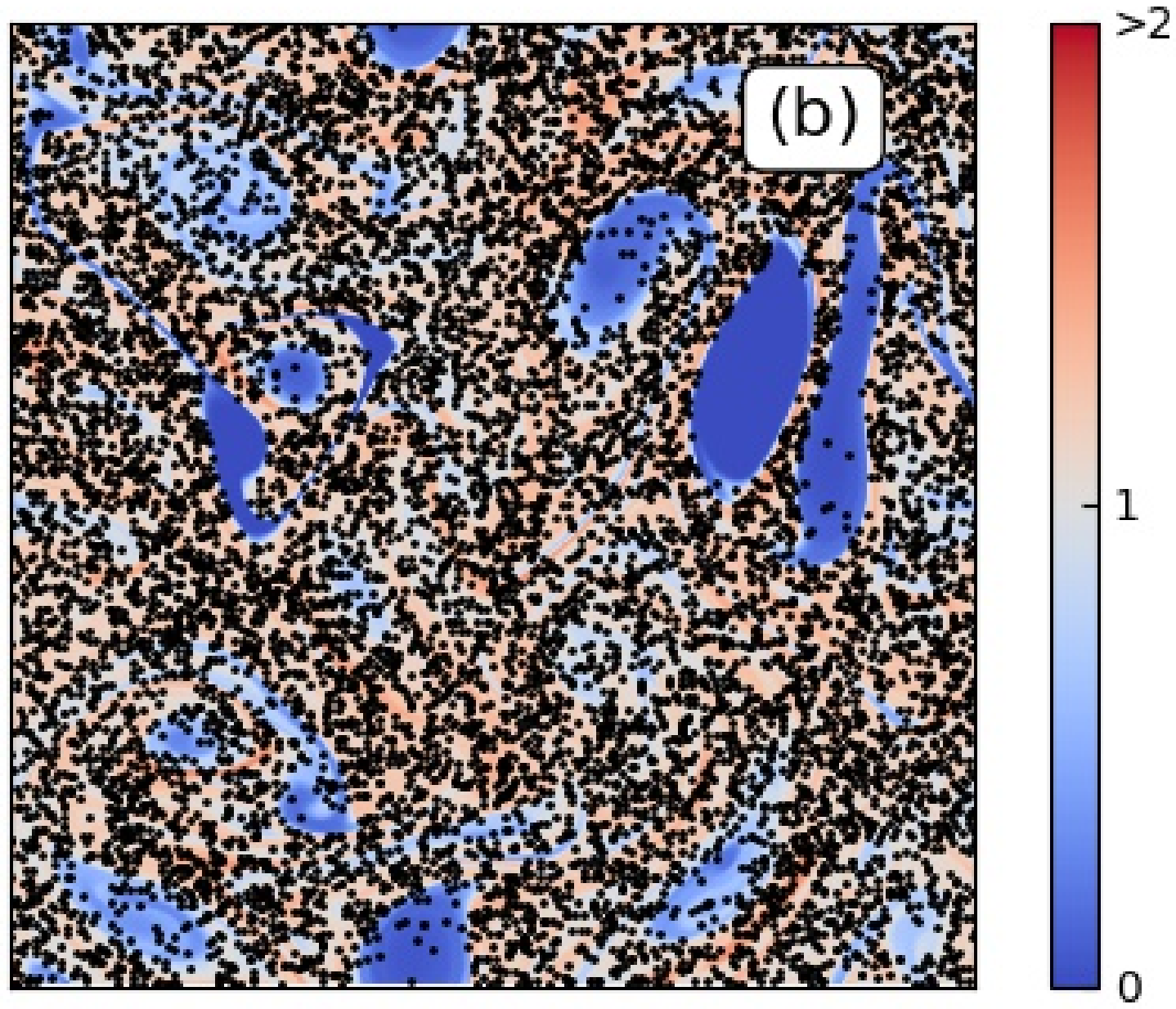} 
\includegraphics[width=0.33\linewidth]{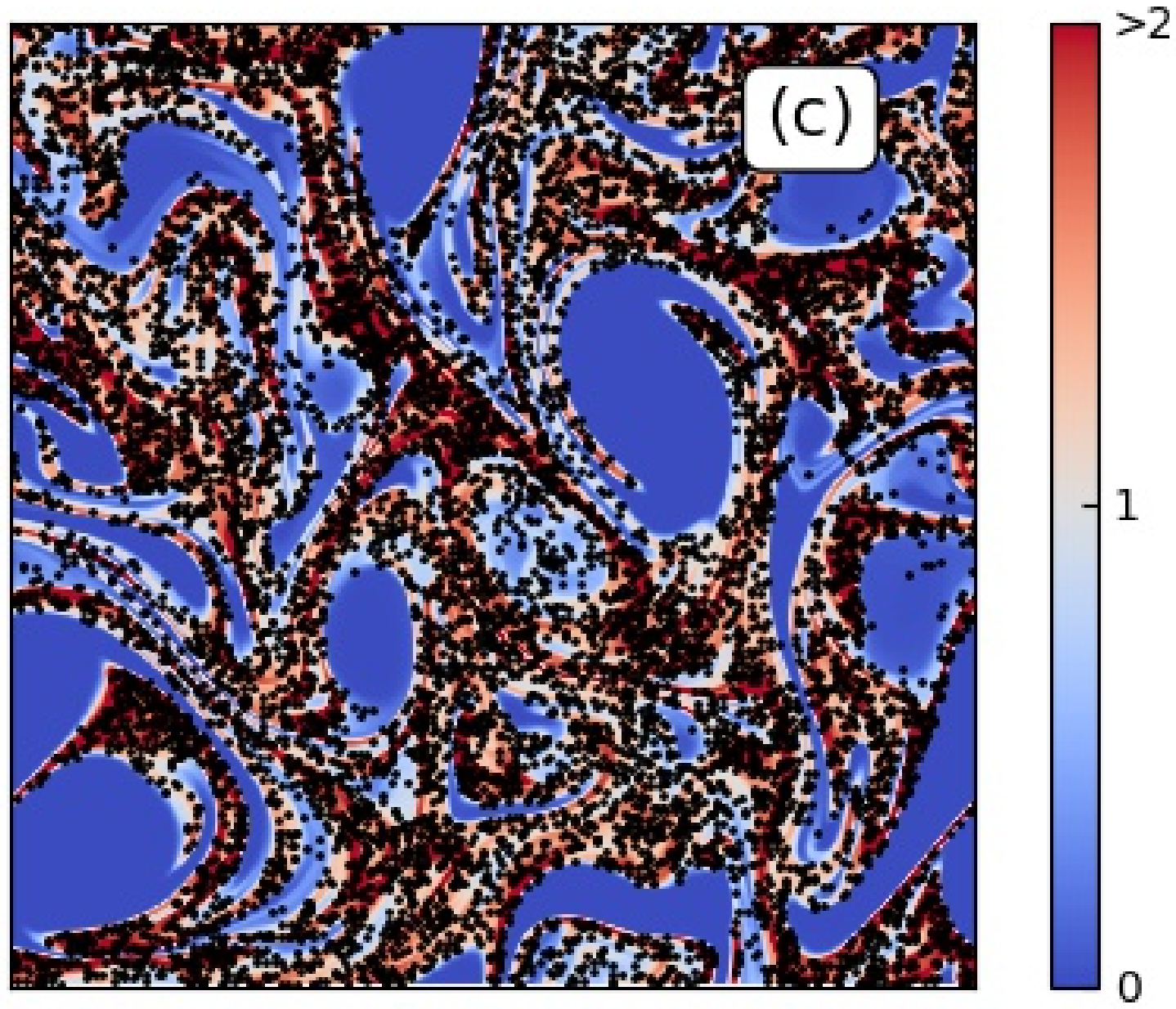}
\caption{\label{fig:snapshot}  Representative snapshot of the (a) gas-vorticity field 
and dust-density field for (b) $\St=1.7\times 10^{-2}$ and (c) $\St=8.6\times 10^{-2}$.
In (a) we have also plotted the compensated time-averaged energy spectrum of gas-velocity. 
The black line indicates the $k^{-4.2}$ scaling.
In (b) and (c) we have overlayed the position of dust particles from our Lagrangian simulations. 
We have used $N=1024$ grid points in each direction in our Eulerian simualtions, and $\Np= 10^4$ 
particles in Lagrangian simulations.  We use $\nu=10^{-5}$, $\alpha=10^{-2}$,
$f_0=5\times 10^{-3}$ and $\kf=4$. This generates a two-dimensional turbulent flow with a 
forward cascade [(a)(inset)] with $\Rey= 1319$ and Kolmogorov time scale $\teta\approx 2.9$ \cite{per11}.  
We choose $\St =3.4\times 10^{-3}, 1.7\times 10^{-2}, 3.4 \times 10^{-2}, 8.6 \times 10^{-2}$, 
and $\St=1.7\times 10^{-1}$. 
}
\end{figure*}
To solve for the gas-velocity, 
we evolve the $2D$ incompressible Navier--Stokes (NS) equations in the 
vorticity-streamfunction formulation:
\begin{subequations}
\begin{align}
\delt \omega + \uu \cdot \grad \omega &= \nu \lap \omega -\alpha \omega + f^{\omega} \/; \\  
\nabla^2 \psi &= \omega \/. 
\end{align}
\label{eq:ns}
\end{subequations}
Here the vorticity $\omega\equiv \curl \uu\cdot\ez$ and 
the streamfuction $\psi$ is defined by 
$\uu\equiv -\curl(\psi\ez)$ where $\uu$ is the gas-velocity, $\nu$ is the gas-viscosity, $\alpha$ is the Ekman drag coefficient,  and we employ the Kolmogorov 
forcing  $f^{\omega}=-f_0 \kf \cos(\kf x)$ that has been used in
earlier experimental \cite{bur99,rivthesis,tith17} and 
numerical \cite{pla91,per09,per11,tsa05} studies to drive the
flow. Here $f_0$ and $\kf$ are the forcing amplitude and 
the wavenumber respectively. To model the dust phase in 
Lagrangian frame we solve  \Eq{eq:HIP}.  To model the dust phase in 
Eulerian frame we solve: 
\begin{subequations}
\begin{align}
\delt \rho + \dive (\rho\vv) & = 0 \/, \label{eq:dustden} \\  
\delt (\rho \vv) + \dive(\rho \vv\otimes\vv) &=  \frac{\rho}{\taup} \left[\uu- \vv \right] \/. \label{eq:dustvel}
\end{align}
\label{eq:enp}
\end{subequations} 
Here $\rho$ is the dust-density field and $\vv$ is the dust-velocity
field and the symbol $\otimes$ denotes direct 
product of two vectors.
We simultaneously numerically integrate Eqs.~\eqref{eq:HIP},
\eqref{eq:ns}, and \eqref{eq:enp} in a square box 
of length $2\pi$ with periodic boundary contions. 
The Eulerian equations, \Eq{eq:ns} and \Eq{eq:enp}, are solved with
$N^2$ collocation points and the 
Lagrangian equations are solved for $\Np$ particles. 
We solve  \Eq{eq:ns} using a pseudo-spectral method~\cite{Can88}. 
To solve the Lagrangian equation, \Eq{eq:HIP}, we need to know the gas-velocity at off-grid points. 
This is acheived by a linear interpolation from neighbouring grid points. 
Because of the presence of sharp-gradients in the density field $\rho$, we employ the 
Kurganov-Tadmor scheme \cite{Centpack,kurt00} to discretize the flux terms in Eq.~\eqref{eq:enp}.
For time integration we use a second-order Adams-Bashforth scheme.
The same code, without the Eulerian description of dust phase, has been used in earlier publications~\cite{per11,per11b}.
\section{Results}
\label{results}
In \Fig{fig:snapshot} we plot  representative snapshots of (a) the gas-vorticity field, and the 
dust-density field for (b) $\St=1.7\times 10^{-2}$ and (c) $\St=8.6\times 10^{-2}$.
We also overplot the instantaneous position of the dust particles in (b) and (c) above. The 
compensated plot of the gas energy spectrum [inset of \subfig{fig:snapshot}{a}] shows a dacade of 
forward cascade regime $E(k)\sim k^{4.2}$.
The different parameters we have used in our simulations can be found in the caption 
of \Fig{fig:snapshot}. 

\subsection{Density accumulation and Particle clustering}
\begin{figure}[!h]
\begin{center}
\includegraphics[width=0.95\linewidth]{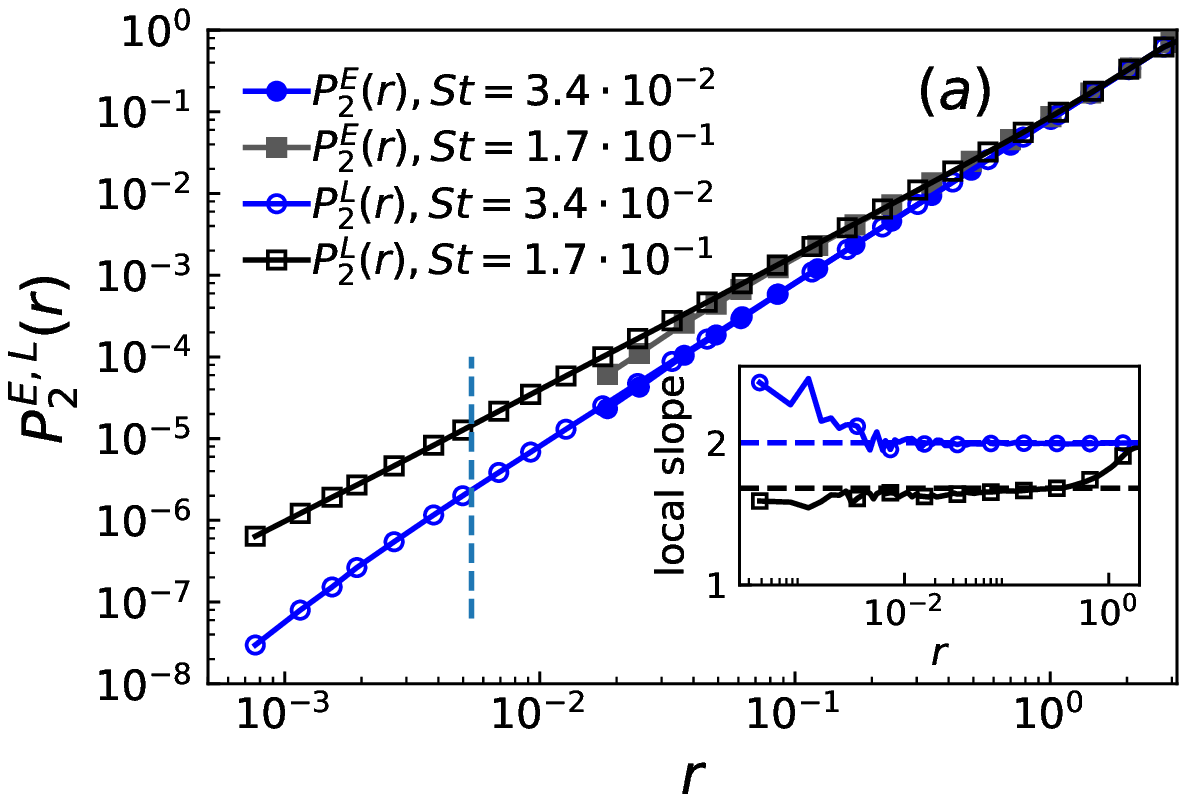} \\
\includegraphics[width=0.95\linewidth]{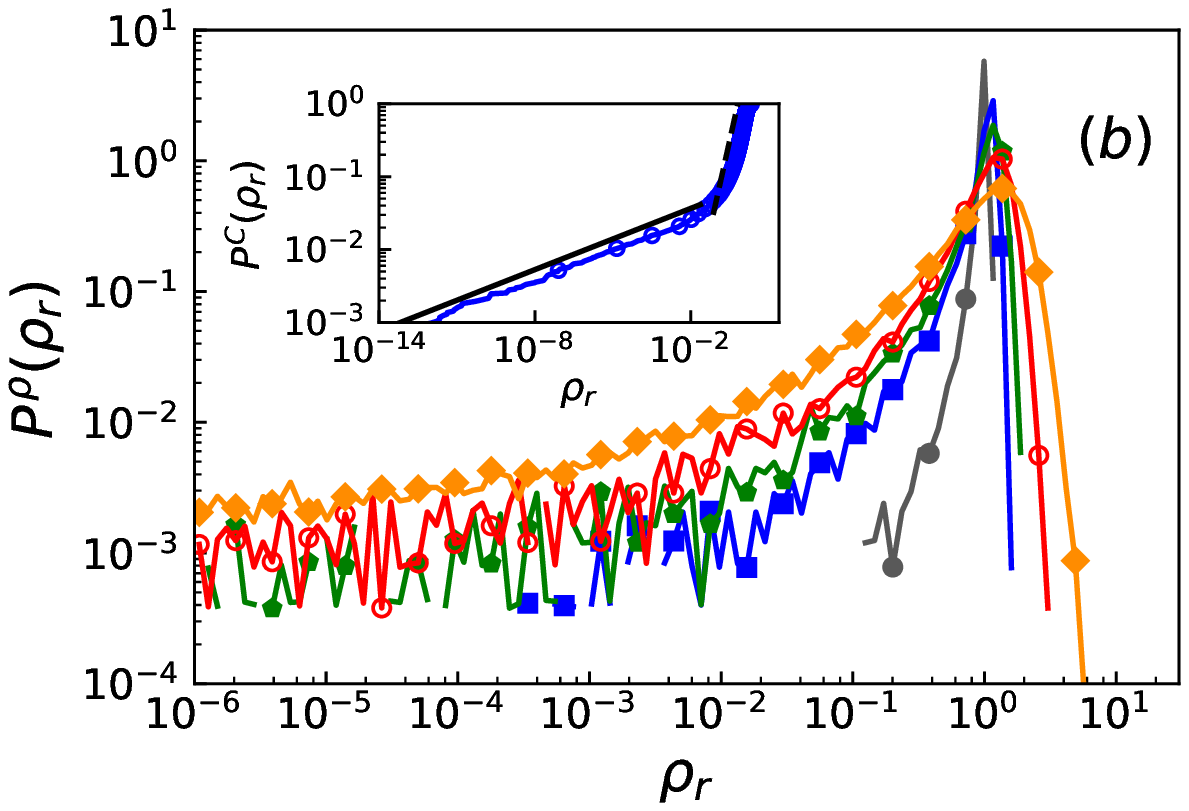} 
\end{center}
\caption{\label{fig:rho_eu} (a) Log-log plot of the dust probability density function (PDF) in 
Eulerian $P_2^{{\rm E,L}}(r)$ for $\St=3.4 \times 10^{-2}$ and $\St=0.017 $. 
(Inset) Local slope  $d_2= d \log{P_2^{\rm L}(r)}/d \log(r)$ analysis  gives  
$d_2=2$ for $\St=3.4\times 10^{-2}$ (blue circle) and $d_2=1.64$ for $\St=0.16$ (black square).
(b) Log-log density--weighted PDF of  the coarse grained dust-density $\rhor$ for different 
values of the Stokes number, $\St=3.4\times 10^{-3}$ (black circle), $\St=0.017$ (blue square), 
$\St=0.034$ (green pentagon), $\St=0.086$ (red empty circle), and $\St=0.17$ (orange diamond). 
(Inset) Cumulative PDF of $\rhor$ for $\St=0.086$ shows that  $\PC(\rhor)\sim \rhor^{1.6}$ for 
$0.1 < \rhor < 1$ (black dashed line)  and $\PC(\rhor)\sim \rhor^{0.15}$ for 
$\rhor<0.1$ (black line). }
\end{figure}

Earlier studies~\cite{druz98,fev05,rani03,rani04,bof07b}, which have tried to compare Lagrangian 
and Eulerian approaches to studying dust-gas multiphase flows, have almost exclusively concentrated 
on comparing clustering of particles in the Lagrangian framework with the enhancement of 
dust-density in the Eulerian framework.  
In \subfig{fig:snapshot}{b} and  \subfig{fig:snapshot}{c} we overlay the instantaneous positions of
dust particles on a pseudocolor plot of a snapshot of  the dust-density field $\rho$ at the same 
time.  In accord with earlier studies \cite{rani03,rani04}, we also find that regions of 
clustering and regions of density accumulation lie on top of each other. 
Furthermore, as expected, we observe that clustering (density accumulation) is enhanced on 
increasing $\St$. 
To make a quantitative comparison, we calculate the probability density function $\PE_2(r)$ 
of the dust-density from our Eulerian simulations, 
\begin{equation}
\PE_2(r) = \left \langle{\int_0^r \rho(0)\rho({\bm r}^\prime) d {\bm r}^\prime}\right \rangle \/,
\label{eq:geu}
\end{equation}
where the symbol $\bra{\cdot}$ denotes averaging of space and time in the statistically 
stationary state of turbulence, and compare it against the probability of finding two particles 
within a distance $r$, $\PL_2(r)$ obtained from our Lagrangian simulations.
In \subfig{fig:rho_eu}{a} we plot, in log-log scale, $\PE_2(r)$ and $\PL_2(r)$, 
for two representative values of $\St$ and in the inset we plot the corresponding local slopes.  
The Eulerian and the Lagrangian data agrees very well with each other.   
From the scaling behavior of $\PE_2(r)\sim r^{d_2}$ as $r \to 0$, we find that the exponent 
$d_2\approx 2$ for $\St=0.034$ and $d_2\approx1.64$ for $St=0.17$.
Similar comparison was earlier made in Refs.~\cite{rani03,rani04}  
in three-dimensions using the ``equilibirum--Euler'' 
approximation~\footnote{ Within the equilibrium--Euler approximation one assumes 
that $\taup$ is small hence the dust velocity is approximated to be 
$\vv \equiv \uu - \tau D\uu/Dt\/$.}.

In \subfig{fig:rho_eu}{b} we plot the density weighted PDF $\Pr(\rhor)$ of the 
dust density $\rho$ coarse-grained over a scale, $r\equiv L/16$, which is inside the inertial 
range. Henceforth the superscript $\rho$ would denote density-weighing. 
The Lagrangian analog of $\Pr$ is the quasi-Lagrangian particle number-density 
(averaged over a box of size $r$) which was calculated -- for 3D turbulent flows -- 
in Ref.~\cite{bec07}. Similar to Ref.~\cite{bec07} we find that 
$\Pr\sim \delta(\rhor)$ for small $\St$ whereas, for large $\St$ the PDF broadens and a 
power-law tail seem to appear as the left tail near $\rhor\sim 1$. 
To investigate the left-tail in greater details, we plot in the inset of 
\subfig{fig:rho_eu}{b} the cumulative probability distribution function $\PC(\rhor)$ which 
is should scale in the same way as the $\Pr(\rhor)$. 
We find two power-law regions: for very small $\rhor<10^{-1}$, 
$\PC(\rhor)\sim \rhor^{0.15}$ whereas for intermediate 
$10^{-1}<\rhor<1$, 
$P^C(\rhor)\sim \rhor^{1.6}$. 
The power-law behavior for intermediate $\rhor$ is consitent with the findings of 
Ref.~\cite{bec07}. 
The behavior at very small $\rhor$ is consistent with presence of 
voids (regions with low density).
\subsection{Spectrum of dust-density and dust-velocity}
\begin{figure}[!h]
\begin{center}
\includegraphics[width=0.95\linewidth]{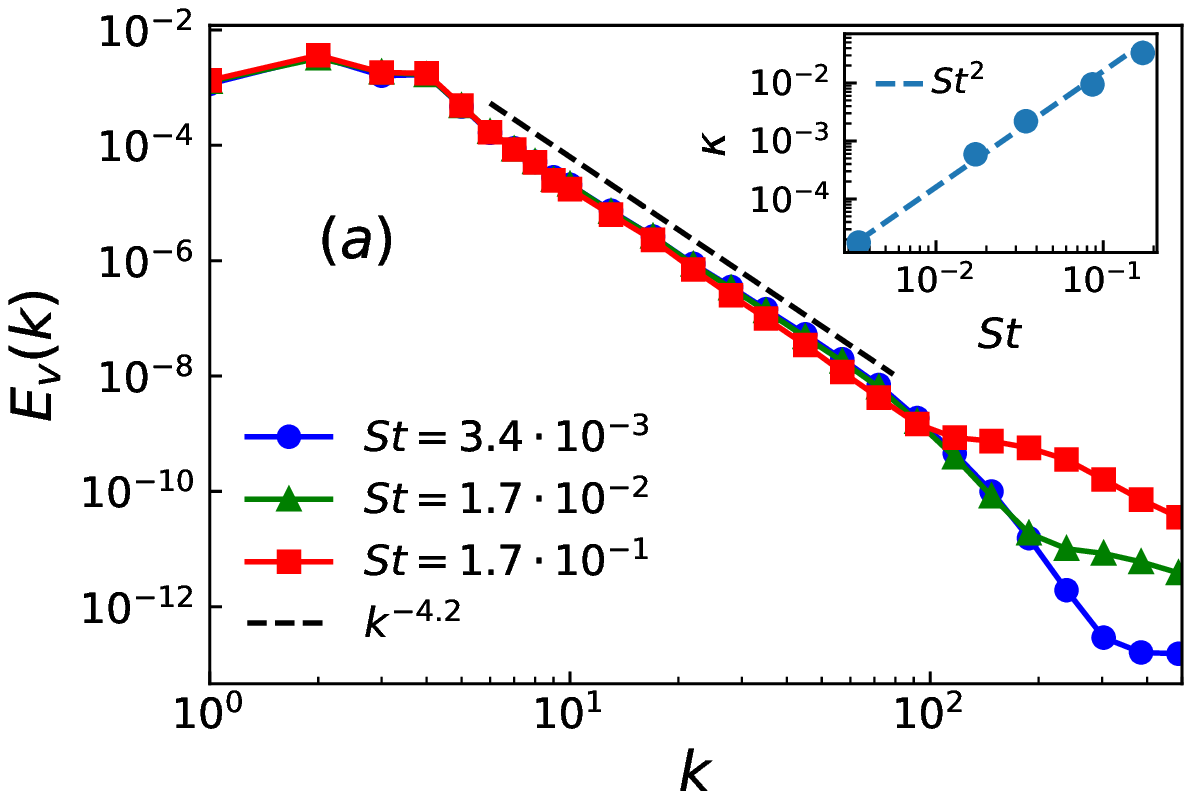} \\ 
\includegraphics[width=0.95\linewidth]{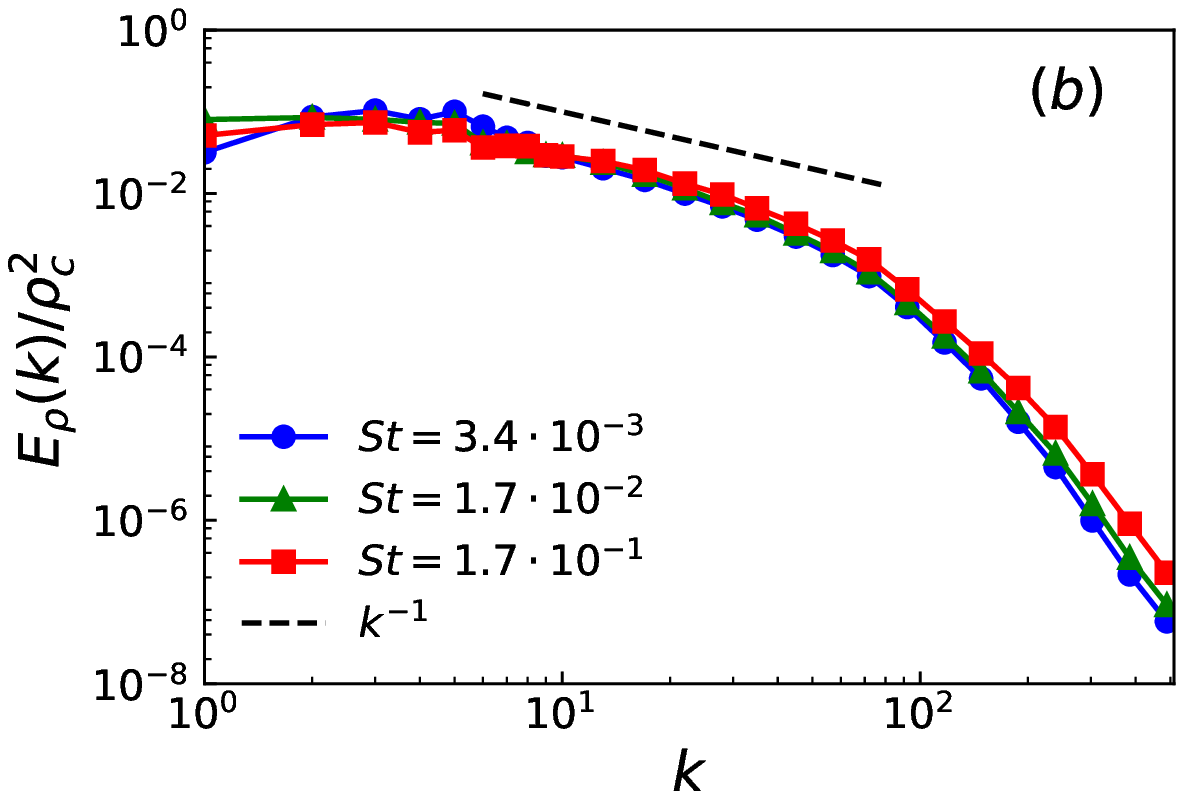}
\end{center}
\caption{\label{fig:spectrum} Log-log plot of the spectrum of (a)
  dust-velocity and  (b) dust-velocity for three values of $\St$. In
  (a) the black line show a slope of $-4.2$ 
which is also the slope of the spectrum of gas-velocity. 
In (b) the black line shows a slope of $-1$.  }
\end{figure}

Next, we plot the angle-averaged spectrum of dust-velocity and dust-density, $\Ev(k)$ and 
$\Erho(k)$ respectively, as a function of the Fourier mode $k$,  in \Fig{fig:spectrum}. 
The spectrum of dust-velocity follows the spectrum of gas-velocity except at very high Fourier 
modes. 
This is indeed what we expect because the dust-velocity essentailly relaxes to the local 
gas-velocity except at high Fourier modes (small scales) where large gradients appear. 
The spectrum of dust-density shows a power-law with an exponent of $-1$, i.e., $\Erho(k) \sim 1/k$. 
This is similar to the Batchelor scaling observed for the case of passive-scalar advected by 
a chaotic flow.
To the best of our knowledge the dust-density spectrum has not been calculated in 2D turbulence before.
The compressibility of the dust flow can be quantified by calculating,
\begin{equation}
\kappa \equiv \frac{\mid\dive \vv\mid^2}{\mid \nabla \vv\mid^2}\/,
\label{eq:compres}
\end{equation}
which we find, consistent with \cite{bof07b}, to be proportional to $\St^2$, as shown in the inset of \Fig{fig:spectrum}.
\subsection{Topological properties}
\begin{figure}[!h]
\begin{center}
\includegraphics[width=0.8\linewidth]{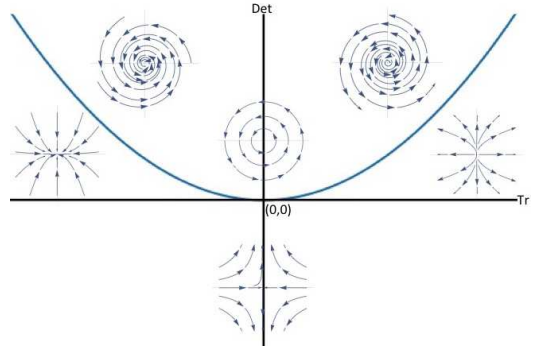}
\end{center}
\caption{\label{fig:mtop} A sketch of possible topological structures in the
two-dimensional in the $\Det$--$\Tr$ plane of the gradient-matrix of the flow. 
Above the horizontal axis ($\Det >0$), the structures are classified into: (Left-Right) Sink, Inward spiral, Center, Outward spiral, and Source. Below the horizontal axis ($Det <0$) only saddles are possible. The parabola $(\Det)^2 = \Tr/4$ (blue line) marks the boundary 
between the vortical and non-vortical regions.}
\end{figure}
The topological properties of any two-dimensional flow at a point is characterised by the
two invariants of the $2\times2$ gradient-matrix, the trace~($\Tr$) and the 
determinant~($\Det$). Four different types of topological structures are possible.
Following Ref.~\cite{Hir04} we sketch them in \Fig{fig:mtop}. 
Along the vertical axis the trace is zero, hence two possible structures are 
either elliptic (center, $\Det >0$ ) or hyperbolic (saddle, $\Det < 0$). 
The topological structures of $\BB$ lies on this line as the gas flow is incompressible. 
 In the more general case of $\AA$ other topological structures are possible. 
Above the parabola with equation, $(\Det)^2= \Tr/4$ lies vortical structures. They 
can be either inward spirals ($\Tr < 0$) or outward spirals ($\Tr > 0$). 
Below the parabola lies the strain-dominated structures, saddles. 
They can be either converging ($\Tr < 0$ ) or diverging ($\Tr > 0$).    

In our case the gradient matrices are $\AA$, and $\BB$, whose components are defined by 
\begin{equation}
A_{ij} \equiv \partial_j v_j \/, {\rm and~} B_{ij} \equiv \partial_j u_i \/.  
\label{eq:AB}
\end{equation}
They are random matrices which are not necessarily Gaussian or white-in-time. 
Clearly, their topological properties are necessarily statistical in nature. 
For $\BB$, the trace is zero due to incompressibility hence its statistical properties are 
completely described by $\Lambda \equiv \Det (\BB)$. 
The statistical properties of  $\BB$ as seen by the dust particles determines statistical 
properties of $\AA$ through \Eq{eq:adot}.
Let us first calculate the statistical properties of $\Lambda$ as seen by dust particles. 

\subsubsection{Topology of the gas velocity field }
\begin{figure}[!h]
\begin{center}
\includegraphics[width=0.95\linewidth]{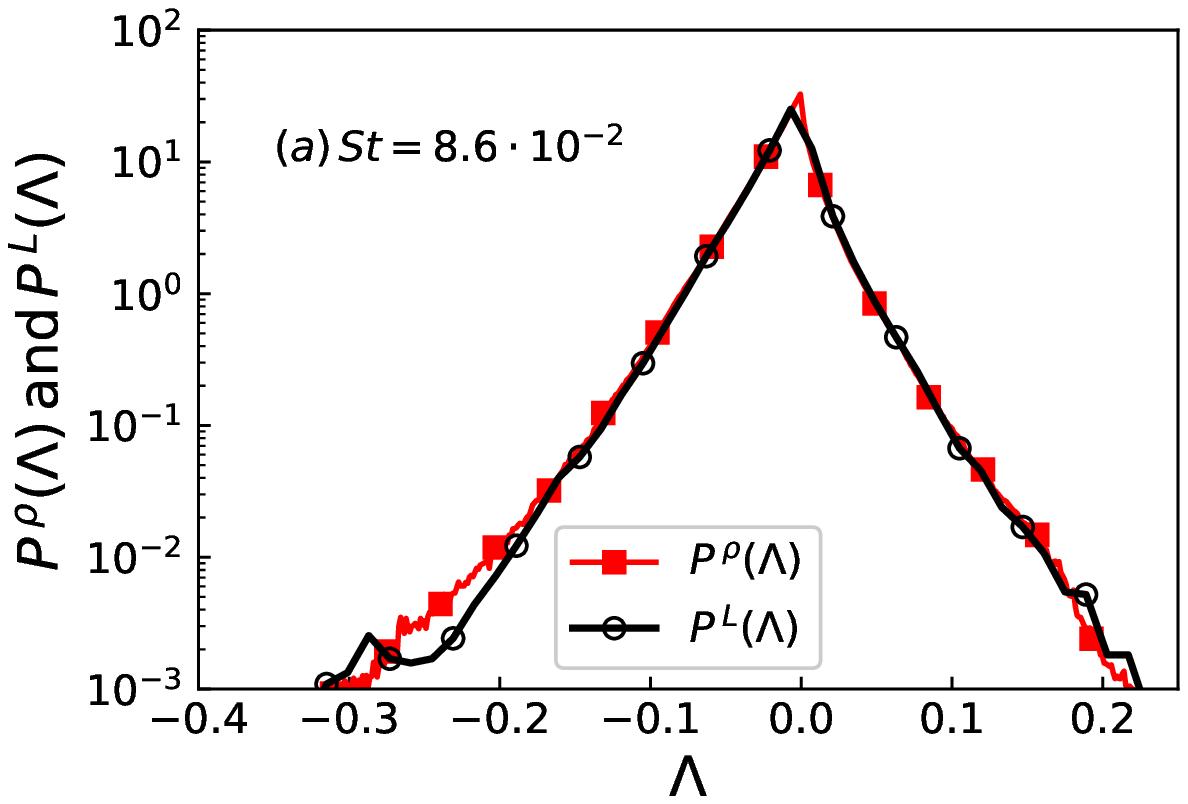} \\
\includegraphics[width=0.95\linewidth]{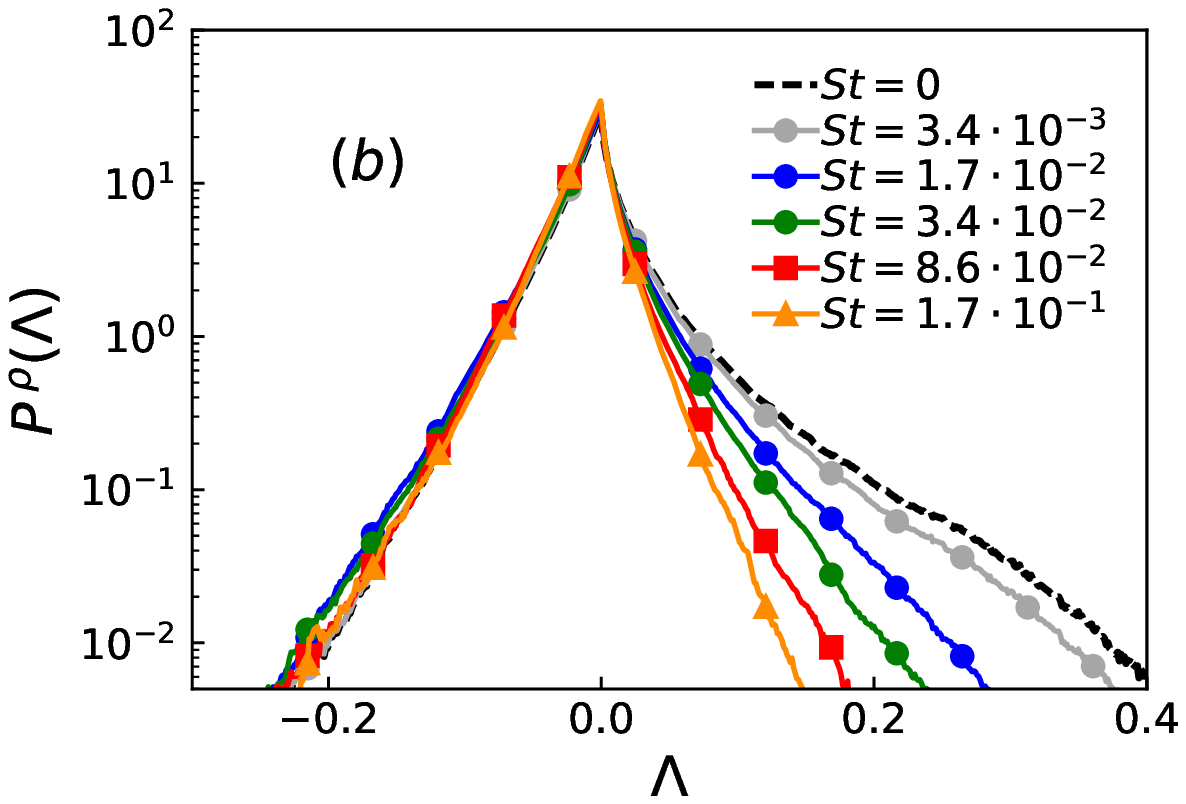}
\end{center}
\caption{\label{fig:lambda} (a) The dust-density weighed PDF of $\Lambda$ calculated from 
Eulerian simulations, $\Pr(\Lambda)$, and the Lagrangian PDF of
$\Lambda$, $\PL(\Lambda)$; $\Lambda \equiv \DB$. 
(b) The dust-density weighed PDF of $\Lambda$, $\Pr(\Lambda)$
calculated from Eulerian simulations 
for several different values of \St. %
}
\end{figure}

In Lagrangian simulations the PDF of $\Lambda$ as seen by the dust, $\PL(\Lambda)$, can be 
directly calculated at the off-grid location of the Lagrangian dust particles using
interpolation. This PDF is plotted in \subfig{fig:lambda}{a} for a 
representative value of $\St=8.6\times 10^{-2}$. 
To calculate the same quantity in the Eulerian case it is not sufficient to
calculate $\Lambda$ at Eulerian grid points but additionally, to account for 
the inhomogeneities in the dust-density, we must use a density--weighted 
PDF $\PEd(\Lambda)$, which is also plotted in \subfig{fig:lambda}{a}. 
We conclude from \subfig{fig:lambda}{a} that these two PDFs agree well with each other. 
This establishes the correspondence between Eulerian and Lagrangian statistics 
that we shall use throughout this paper: the Lagrangian statistics of any quantity 
is obtained by calculating the Eulerian statistics of the same quantity weighed by the 
dust-density. 
Note that this equivalence -- in whose support we shall present more results below -- 
is established for time independent quantities only, it does not imply that a 
similar correspondence can be made between the correlation time of a certain quantity in 
Eulerian and Lagrangian frame. 
Earlier experiments and numerical simulations \cite{riv01,per09,per11} 
have investigated the distribution of vortical and strain dominated
regions for the case of tracers ($\St=0$). 
These studies show that the distribution is positively 
skewed indicating the prominence of vortical regions in the gas flow. 
Since dust particles expelled out of vortical regions we expect the right 
tail of the PDF to decrease faster as $\St$ increases.
This is indeed confirmed in \subfig{fig:lambda}{b}. 
\subsection{Topology of the dust velocity field}
To characterise the topological properties of the dust velocity field we must
use both the determinant and trace of the gradient-matrix $\AA$.
This can be done in a Lagrangian simulation only if for each individual Lagrangian dust particle 
we solve for not only its position and momentum but also for the nine independent 
components, $A_{ij}$ by solving \Eq{eq:adot}. So far, such a calculation has been
done only in Ref~\cite{falkovich2007sling} in three-dimensional turbulent flows. 
Let us start by calculating the topological properties of $\AA$ first in
the Eulerian framework. 

\subsubsection{Topology of dust velocity in Eulerian simulations}
\begin{figure*}
\begin{center}
\includegraphics[width=0.33\linewidth]{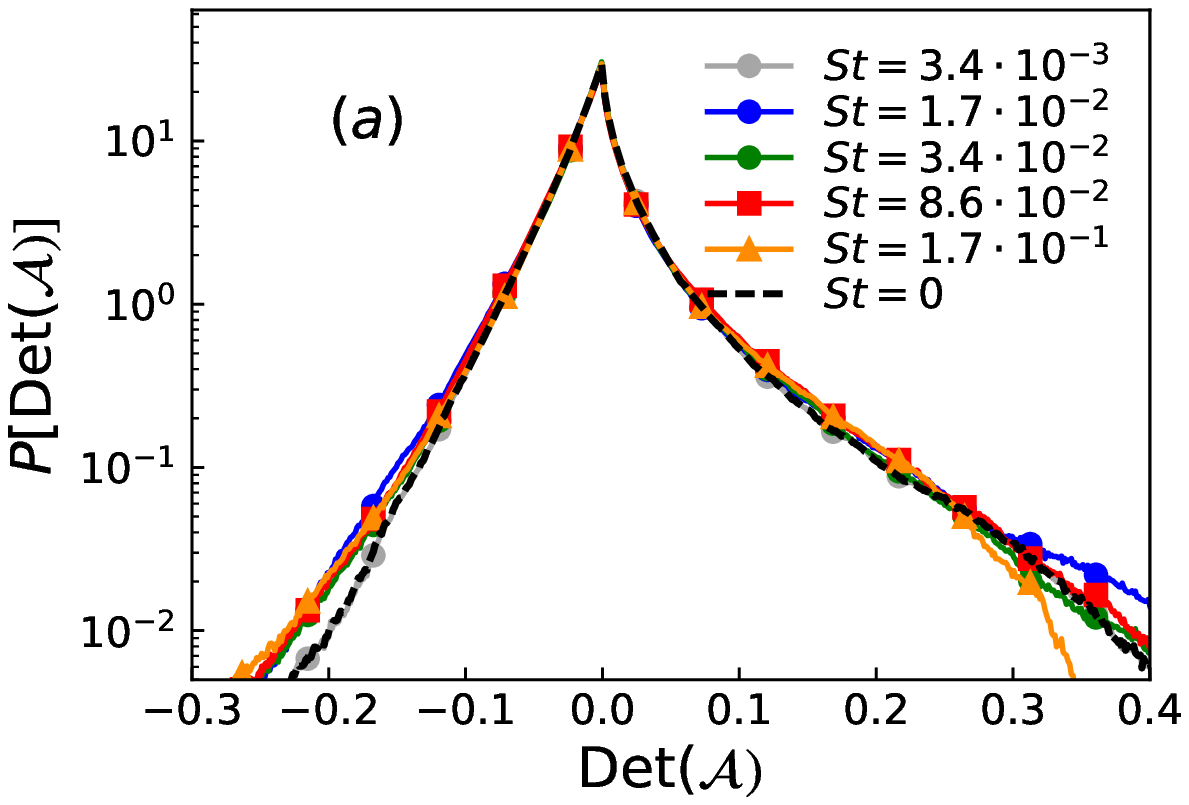} 
\includegraphics[width=0.33\linewidth]{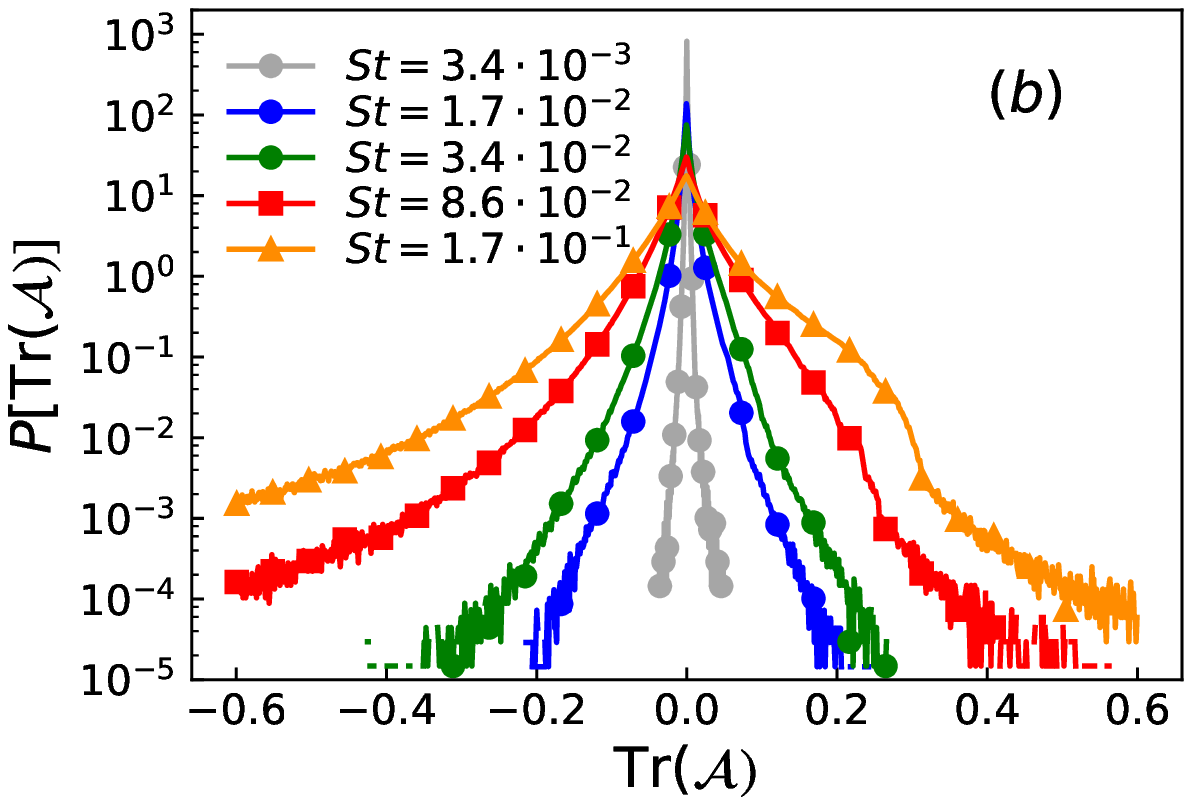} 
\includegraphics[width=0.33\linewidth]{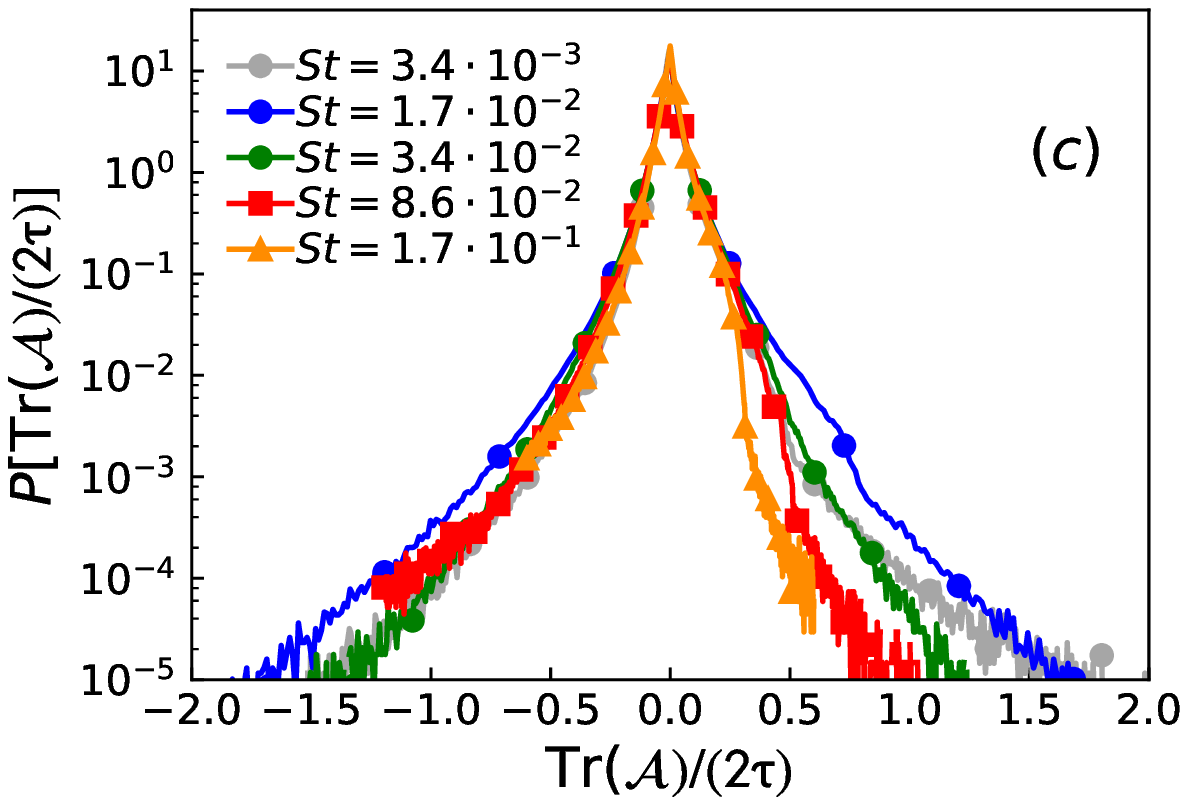}
\end{center}
\caption{\label{fig:topo1}  (a) Probability distribution $P[\DA]$ versus $\AA$ for different Stokes numbers. Note 
that $P[\DA]$ does not change appreciably on changing $\St$; (b) $P[\TA]$ versus 
$\TA$ for different values of $\St$. 
(c) $P[\TA/2 \taup]$ versus $\TA /2\taup$ 
overlap for different values of Stokes number, $\St$.}
\end{figure*}

\begin{figure*}[!h]
\begin{center}
\includegraphics[width=0.33\linewidth]{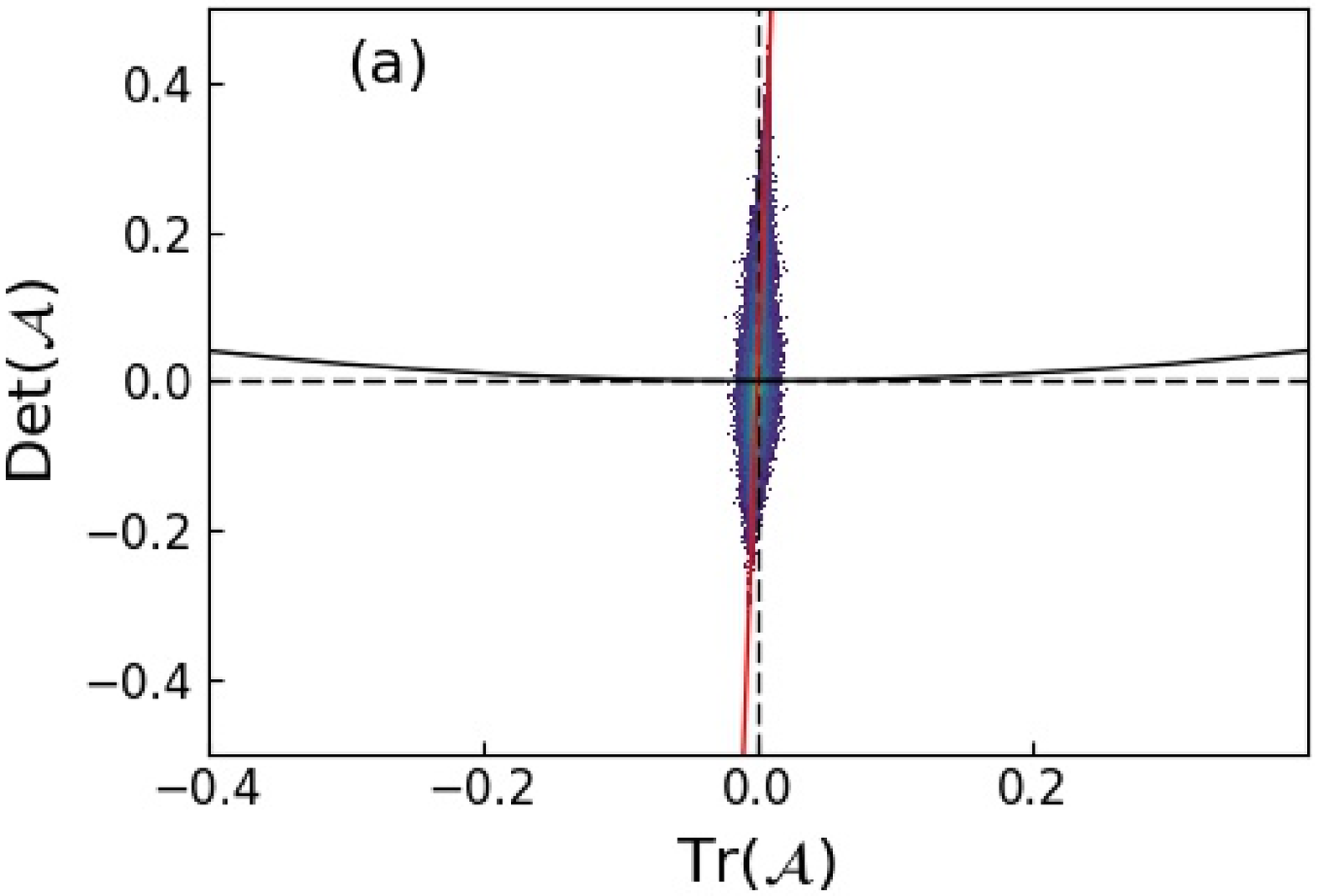} 
\includegraphics[width=0.33\linewidth]{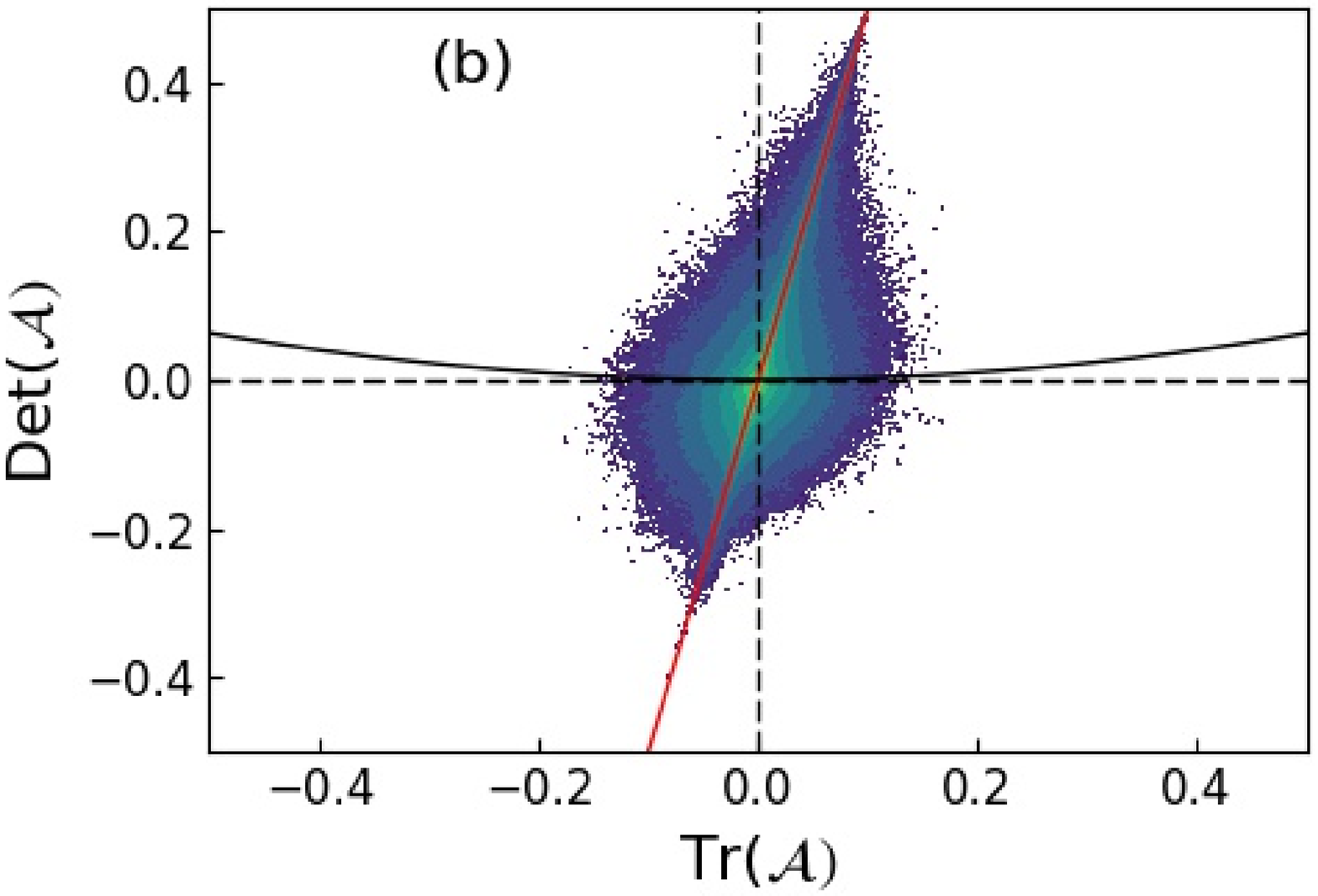} 
\includegraphics[width=0.33\linewidth]{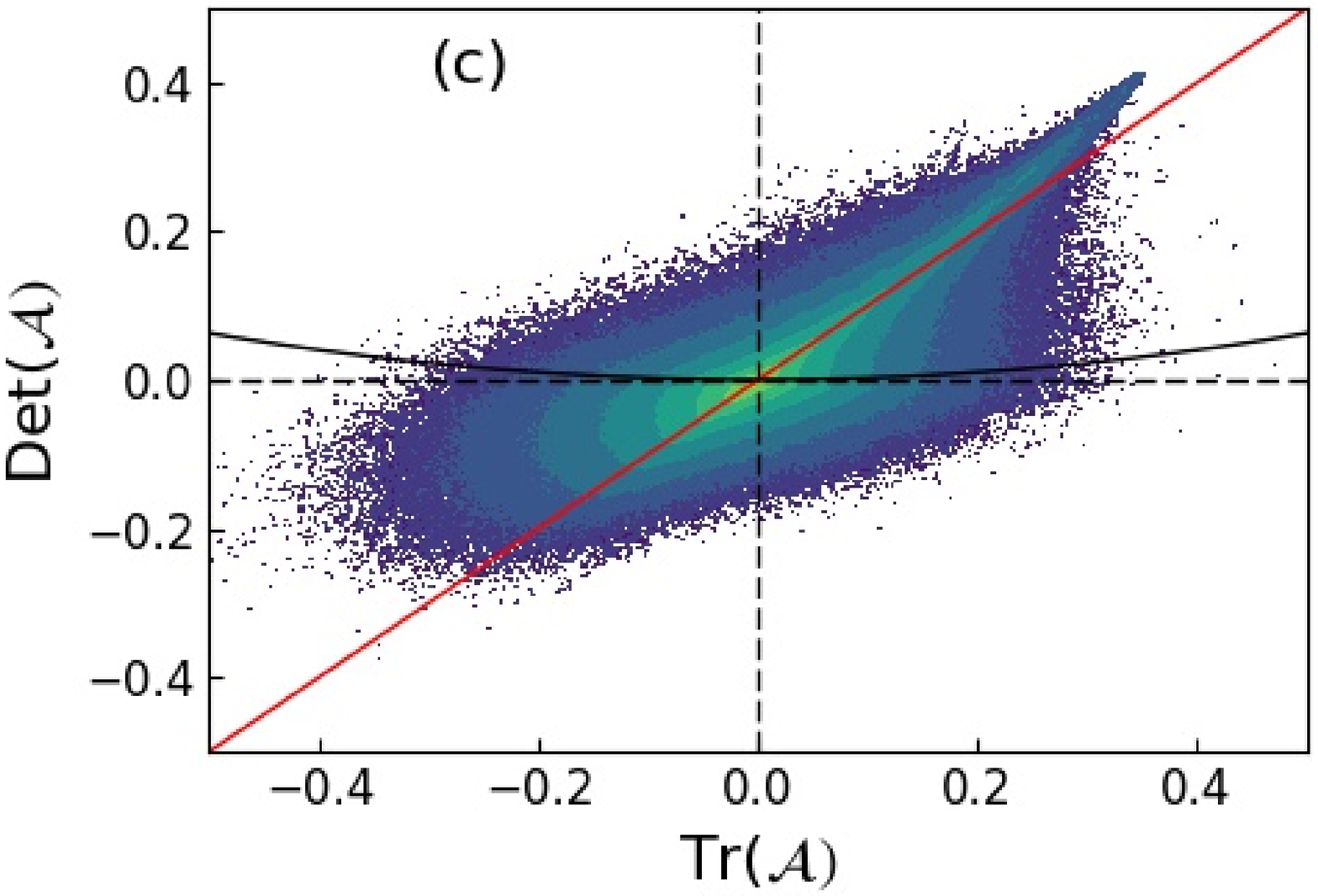}
\caption{\label{fig:topo2} Joint PDF $Q[\TA,\DA]$ for $\St=3.4\times 10^{-3}$ (a), 
$\St=0.017$ (b), and $\St=0.17$ (c). The red line is the equation $\DA=\TA /2\taup$. }
\end{center}
\end{figure*}

In \subfig{fig:topo1}{a} and \subfig{fig:topo1}{b} we plot respectively
the PDF of $\DA$ and $\TA$ for several different values of $\St$.
We find that $P[\DA]$ depends very weakly on $\St$; 
whereas $P[\TA]$ is negatively skewed and the distribution 
broadens as $\St$ increases. 
Both these results can be understood by noting that for small $\taup$, the dust velocity $\vv$ 
can be expanded in powers of $\taup$ as  
$\vv=\uu-\taup D{\uu}/Dt + O(\taup^2)$. 
It is then easy to show that  to the leading order $\DA\approx \DB$ 
[note that $\DA=\DB$ for $\taup=0$] and 
$\TA= 2\taup\DB$. 
Thus, we expect that $P[\DA]=P(\Lambda)$ for small $\St$, which is indeed 
confirmed in \subfig{fig:topo1}{a}. 
On rescaling $\TA$ by $2\taup$ we find a reasonable 
collapse for small $\St$ in \subfig{fig:topo1}{c}. 
The distributions calculated above are projections of the joint probability distribution 
$Q[\TA,\DA]$ on either $\DA=0$ or $\TA=0$  axis and cannot 
distinguish between different types of topological structures in the 
flow (see \Fig{fig:mtop}). To investigate the detailed flow topology, 
we now plot  $Q[\TA,\DA]$ for three different values of $\St$ in \Fig{fig:topo2}. 
For small values of $\St=3.4\times 10^{-3}$, $\TA \approx 0$ and the $Q$ is concentrated 
along the vertical axis.  On increasing the $\St$, the joint pdf becomes broader and tilts clockwise;
this is indeed what we expect because for small $\St$; 
$\DA \approx \TA/(2\taup)$.
In terms of topological structures, this implies that we get more diverging
spirals and converging saddles in the flow and less 
of converging spirals and diverging saddles. 
It is the converging part that contributes to the increase in dust-density
and this comes from the saddles, which are the strain-dominated points. 
As the $\St$ is increases further the joint PDF, $Q$ gets wider 
and it does not any longer satify the simple relation:
$\DA \approx \TA/(2\taup)$, but the general trend 
among the different topological structures remain the same. 

\subsubsection{Topology of density-weighed dust velocity}
\begin{figure*}
\begin{center}
\includegraphics[width=0.33\linewidth]{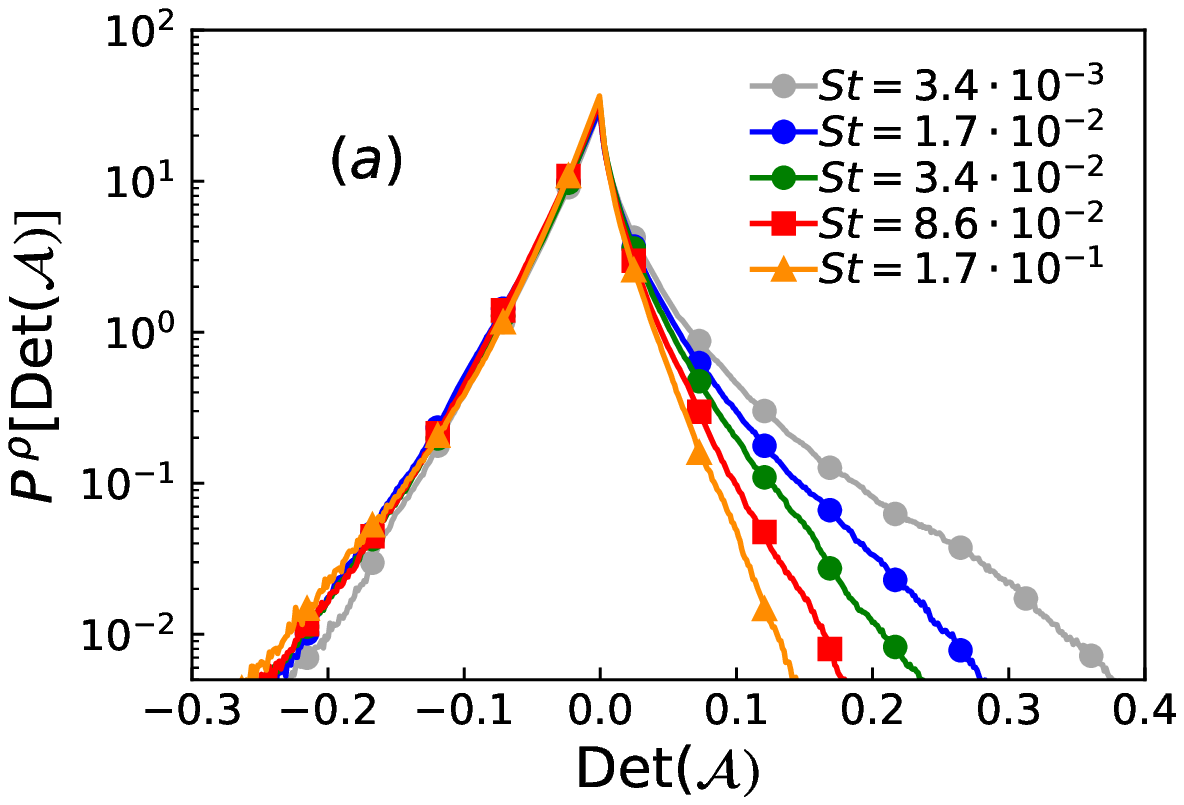} 
\includegraphics[width=0.33\linewidth]{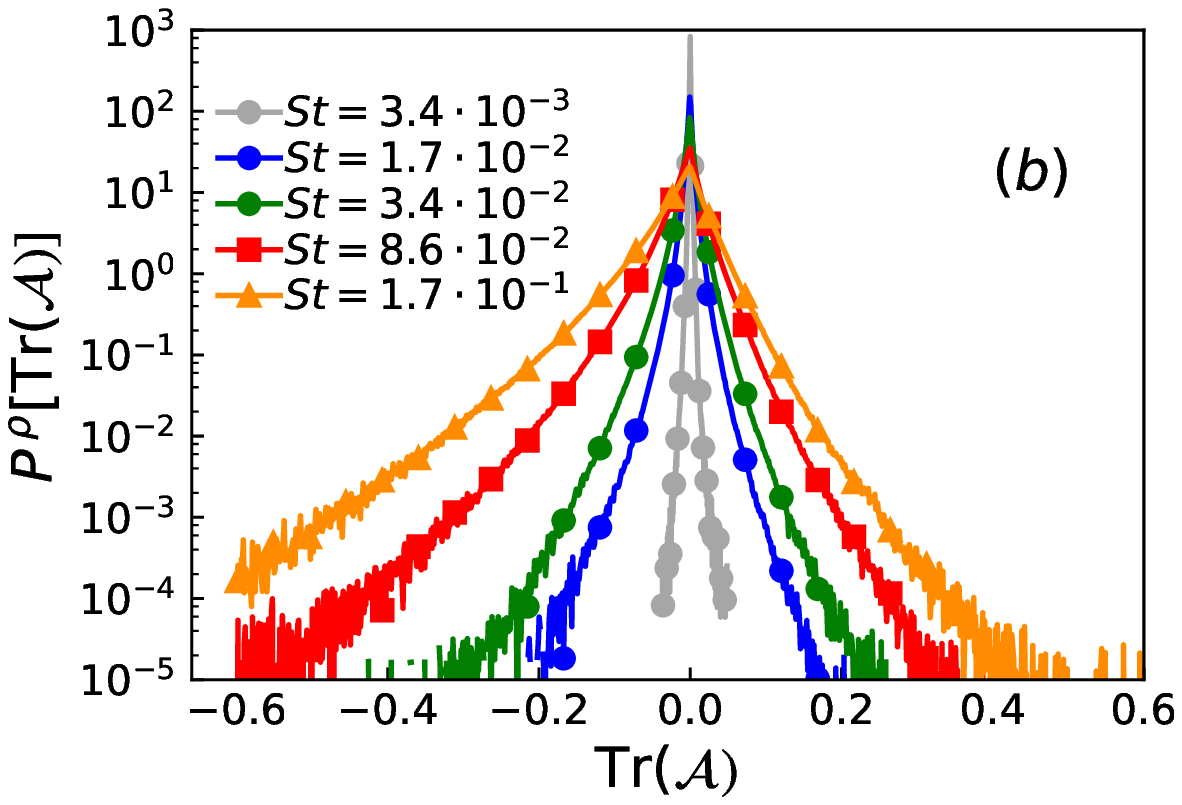} 
\includegraphics[width=0.33\linewidth]{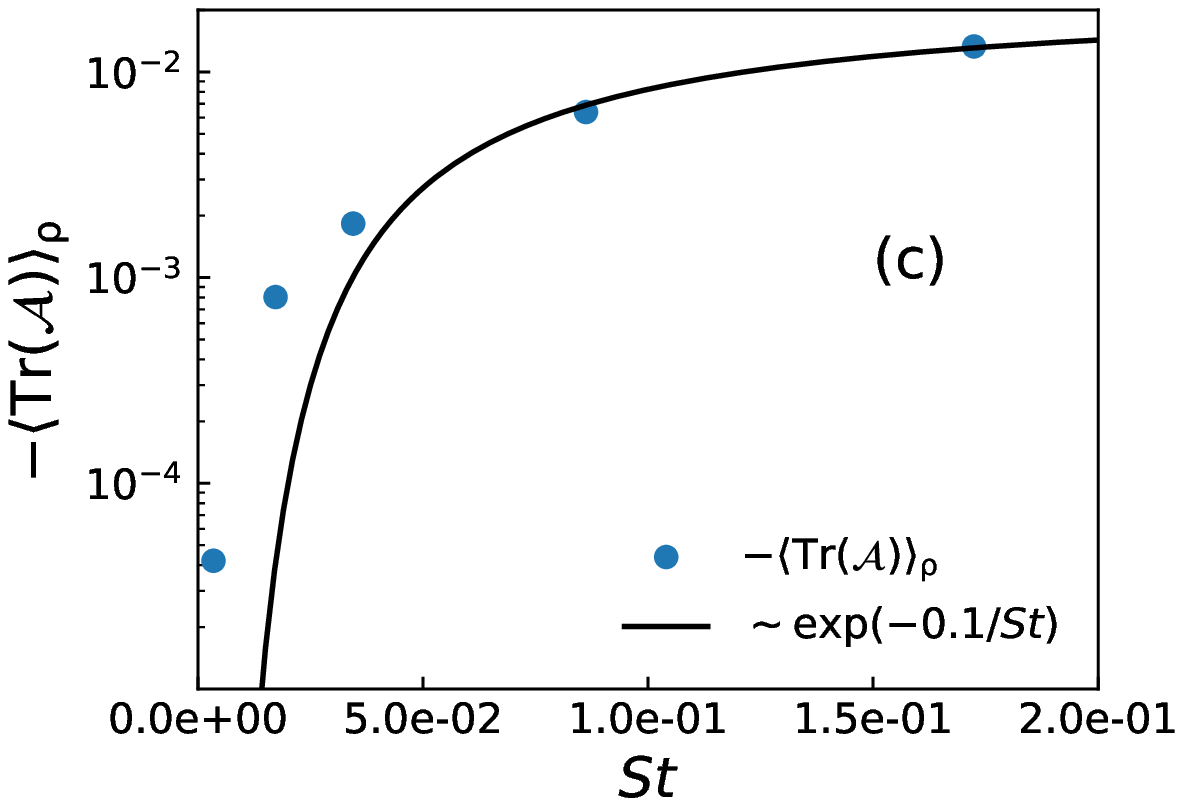}
\end{center}
\caption{\label{fig:topo3} Density-weighted PDFs, $\Pr[\DA]$ (a), and $\Pr[\TA]$ (b) 
for different values of $\St$. (c) The negative of the  density-weighted mean of the
trace of $\AA$,  $-\bra{\TA}^{\rho}$ as a function of $\St$. The continious line in the figure
is the expression $\exp(-0.1/\St)$}
\end{figure*}

We have shown before that the Lagrangian statistical properties of $\Lambda$ can be 
calculated from the Eulerian simulations by calculating the density-weighed statistical properties.
Motivated by this agreement, we now calculate the density-wighted statistical properties of
$\DA$ and $\TA$ from our Eulerian simulations and claim that these are
the same result as would have been obtained by a Lagrangian simulation.  
In \subfig{fig:topo3}{a} and \subfig{fig:topo3}{b} we plot respectively the density--weighed PDF
of the determinant of $\AA$, $\Pr[\DA]$ and the trace of $\AA$, $\Pr[\TA]$ for
several values of $\St$. We find that on increasing $\St$ the right tail of 
$\Pr[\DA]$ systematically decreases. 
This can be understood by first remembering that we have already argued in the earlier section, 
$P[\DA] \approx P[\DB]$ in the Eulerian framework. This argument works equally 
well in the Lagrangian framework too, hence we expect that, 
$\Pr[\DA] \approx \Pr[\Lambda]$.
It has been argued and confirmed from simulations~\cite{bec05} that heavy inertial particles are 
ejected from vortices, hence as $\St$ increases the dust particles sample less and 
less the strain-dominated regions ($\Lambda > 0$), hence the right tail 
of $\Pr[\DA]$ decreases. 

Next we plot the PDF of $\TA$ in \subfig{fig:topo3}{b} for several different
values of $\St$.  This PDF is of particular interest. 
The presence of singularities (caustics) would imply that the trace of $\AA$ would
blow-up. In our simulations we have used a shock-capturing scheme to regularize 
these singularities, hence we are not going to find any blow-up in our simulations.
Although, we may find signature of the singularities in large negative values of
$\TA$. We see in \subfig{fig:topo3}{b} that the PDF of $\TA$ 
has a shallower tail on the negative side than on the positive side. 
We quantify this by calculating the mean value of this PDF, $\bra{\TA}^{\rho}$, 
as a function of $\St$ [see  \subfig{fig:topo3}{c}]. 
Analytical calculations in a simple model~\cite{wilkinson2006caustic} has shown that 
the frequency of caustics increases with $\St$ as $\exp(-C/\St)$ 
with a constant $C$. We find that this expression is provides reasonable 
fit to our data on $\bra{\TA}^{\rho}$  versus $\St$ with $C\approx 0.1$.

\begin{figure*}[!h]
\begin{center}
\includegraphics[width=0.33\linewidth]{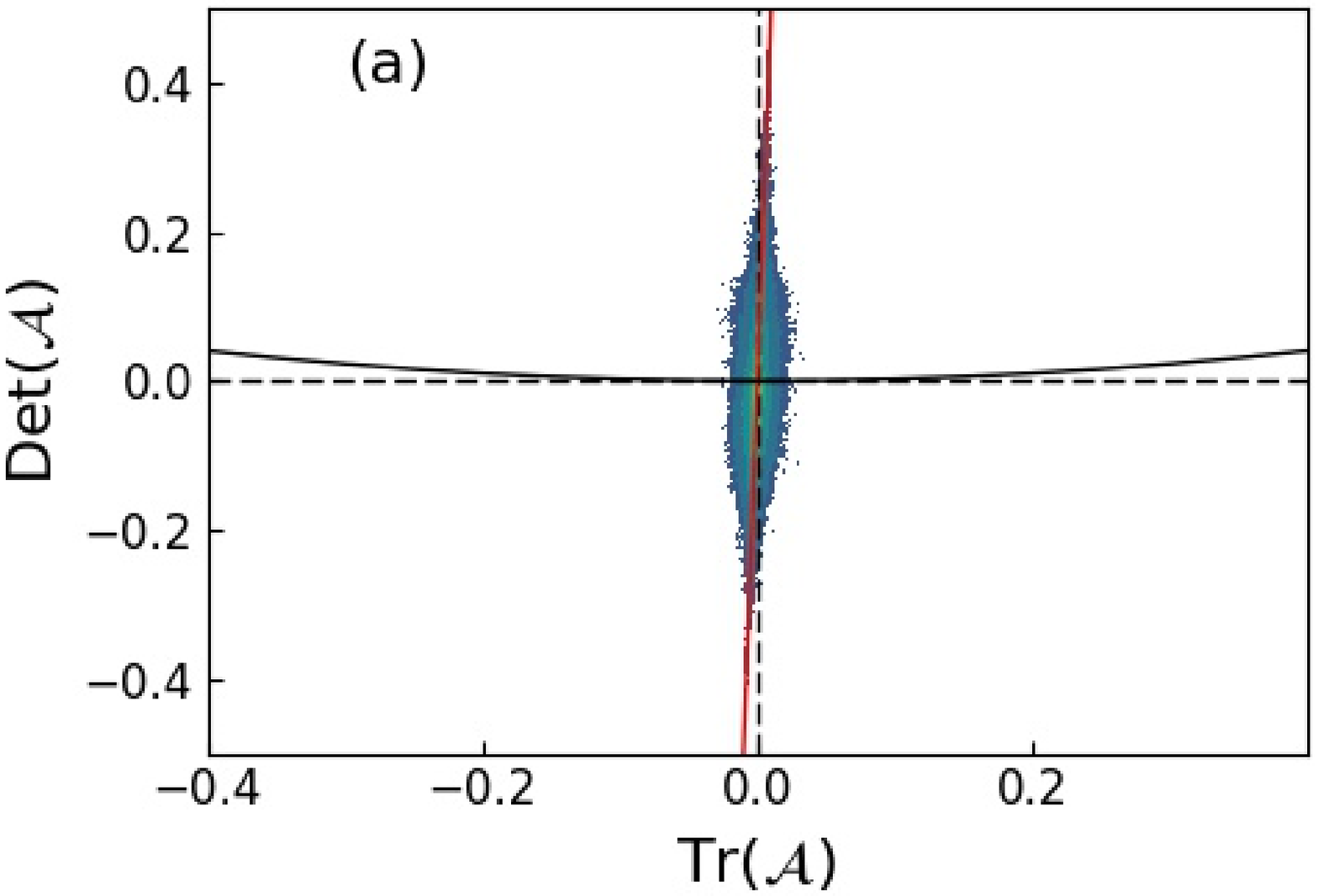} 
\includegraphics[width=0.33\linewidth]{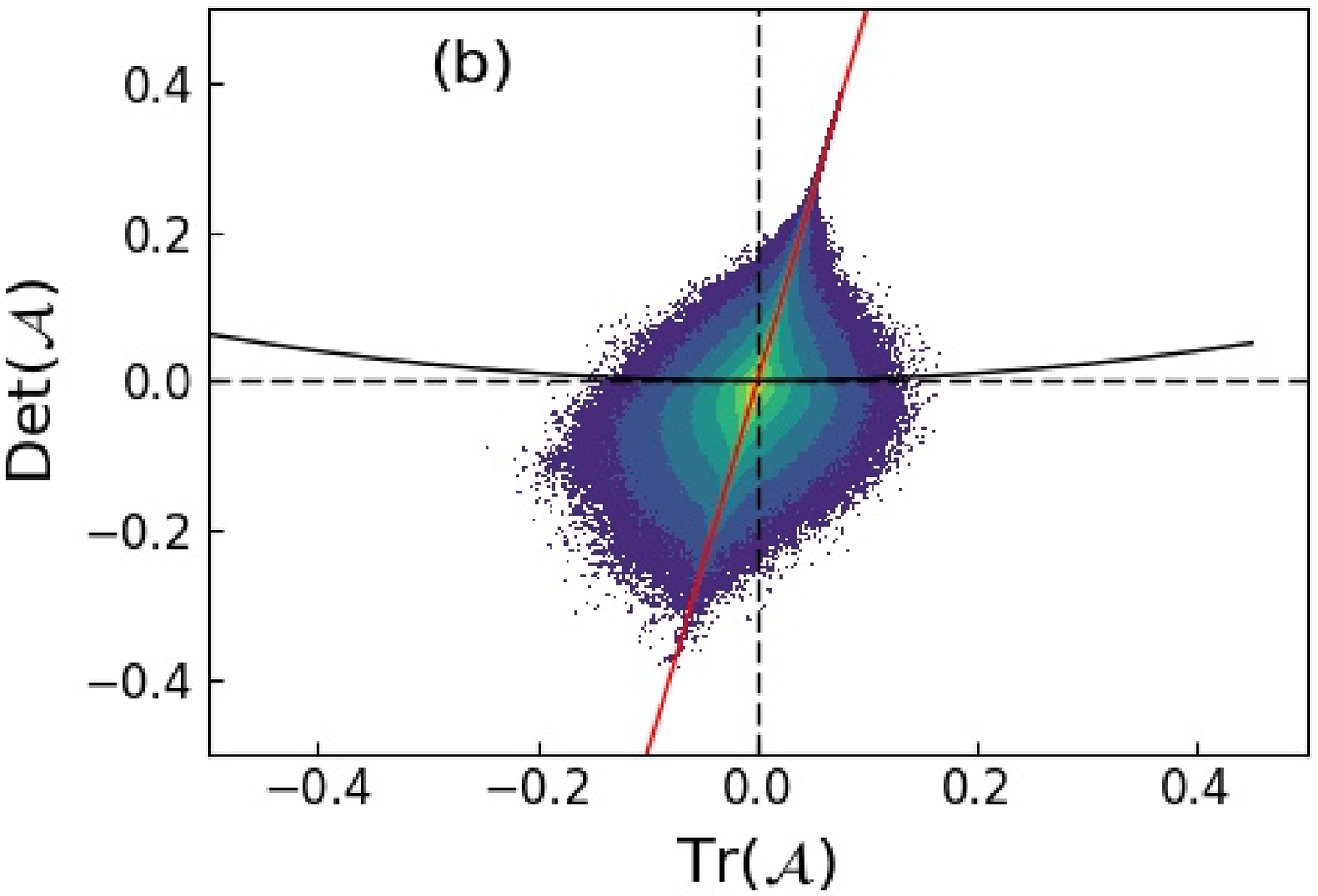} 
\includegraphics[width=0.33\linewidth]{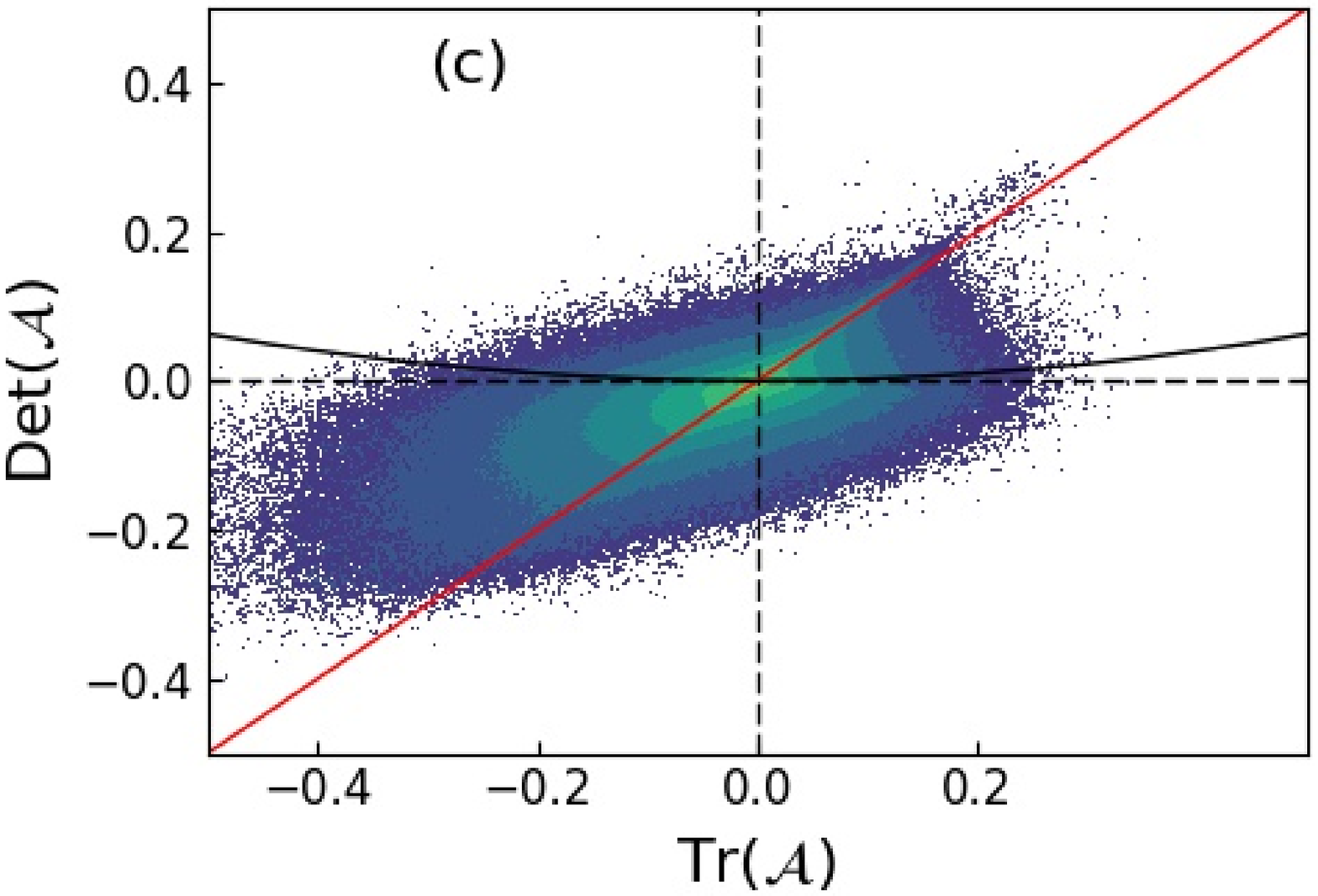}
\caption{\label{fig:topo4} Density- weighted joint PDFs $\Qr[\TA,\DA]$ for 
$\St=3.4\times 10^{-3}$ (a), $\St=0.017$ (b), and $\St=0.17$ (c). 
The red line is the function $\DA=\TA/2\taup$. The wide parabola is the line
$(\DA)^2 = \TA/4$.}
\end{center}
\end{figure*}

Finally, we plot the density-weighted joint probability distribution $\Qr[\TA,\DA]$,
for three different values of $\St$, in \Fig{fig:topo4}. 
By comparing these figures against Eulerian joint PDF,  $Q[\TA,\DA]$ plotted in
\Fig{fig:topo2} we find how the topological structures change between the 
Eulerian and the Lagrangian frames. 
Compared to the Eulerian case, in the Lagrangian (i.e., the density-weighted) case we 
find a qualitative change: there are less spiral structure and more saddles, 
and it is more probable to have converging saddles.  

\section{Conclusion}
\label{conc}
We characterised the topological structures that appear
in homogeneous and isotropic turbulent flows of dust and gas.
The gas flow is assumed to be incompressible but the flow of the dust is 
compressible, hence is more rich is topological structures. 
The topological properties of any two-dimensional flow can be characterised by 
the trace and the determinant of its gradient matrix. 
As we are dealing with a turbulent flow the matrices we deal with are random 
matrices, their topological properties are necessarily statistical. 
Earlier analytical studies~\cite{fal02,wilkinson2006caustic} have suggested that 
the flow of dust develops singularities -- one or more element of the
gradient-matrix, $\AA$ can blow-up --  but the nature of such singularities
in either two or three dimensions has not been elucidated.
Direct numerical simulations necessarily introduces regularisation of such
blow-ups -- our shock-capturing scheme -- hence cannot probe the 
blow-up directly.
Within this limitation, we, for the first time, have characterised the statistics of the 
topological structures found in the dust velocity. 
We have done this in Eulerian framework, and used dust-density-weighted quantities to
translate our Eulerian results to Lagrangian ones.
The singularities of $\AA$ should disappear as $\St \to 0$.
Analytical results in one-dimensional models suggest that
the number of blow-ups should increase as $\exp(-C/\St)$, with a positive constant $C$. 
To the best of our knowledge, Ref.~\cite{falkovich2007sling}, is the only work that
tried to check this result against Lagrangian DNS in three dimensions.
The nature of our simulations do not allow us to directly calculate this quantity, however
we have use the $\bra{\Tr\AA}_{\rho}$ as a proxy of this quantity -- and have 
found reasonable agreement.

A summary of the other results follows:
(a) The dust-density field in our Eulerian simulations have the same correlation dimension 
$d_2$ as obtained from the clustering of particles in the Lagrangian simulations for $\St < 1$. 
(b) The dust-density coarse grained over a scale $r$ in the inertial range, $\rhor$,
calculated from our Eulerian DNS shows large fluctuations. 
We quantify these fluctuations by computing the cumulative probability 
distribution function (CDF), $\PC(\rhor)$. This CDF has a left-tail with power-law fall-off
that indicates presence of voids in dust-density.
(c) The energy spectrum of the dust-velocity has a power-law range with an exponent 
that is same as the gas-velocity spectrum except at very high Fourier modes. 
The spectrum of dust-density also shows a scaling range with an exponent of $-1$.
(d) The compressibility of dust velocity field is proportional to $\St^2$.   
(d) The statistics of topological properties of $\BB$ are the same in Eulerian and Lagrangian 
frames only if the Eulerian data are weighed by the dust-density. 
We use this correspondence to calculate statistics of topological properties of
$\AA$ in the Lagrangian frame from our Eulerian simulations by 
calculating density-weighed averages probability density functions.
In particular, we find that: 
(e) The mean value of $\Tr\AA$ in the Lagrangian frame, $\bra{\Tr\AA}^{\rho}$, is negative 
and its magnitude increases with $\St$ approximately as $\exp(-C/\St)$ with a 
constant $C\approx 0.1$.
(b) For small $\taup$,  $\Det\AA \approx \Det\BB$ and  $\Tr\AA\approx 2\taup\Det\BB$. 
(f) The mean value of the PDF density-weighed $\Det\AA$, $\bra{\Tr\AA}^{\rho}$
is negative and its magnitude increases with $\St$ approximately as $\exp(-C/\St)$
with a constant $C\approx 0.1$. 
(g) The statistical distribution of different topological structures that appear in the
dust flow are different in Eulerian and density-weighed Eulerian cases particularly for
$\St$ close to unity. In both of these cases, for small $\St$ the topological structures 
have close to zero divergence and are either vortical (elliptic) or 
strain-dominated (hyperbolic,saddle). 
As $\St$ increases, the contribution to negative divergence comes mostly from 
saddles and the contribution to positive divergence comes from vortices 
and saddles. Compared to the Eulerian case, the density-weighed Eulerian case has
less inward spirals and more converging saddles. Outgoing spirals are the least 
probable topological structures in both cases.

\section{Acknowledgement}
DM is supported by the grant Bottlenecks
for particle growth in turbulent aerosols from
the Knut and Alice Wallenberg Foundation (Dnr. KAW
2014.0048). DM thanks Akshay Bhatnagar for useful discussions. 

%

\end{document}